\documentstyle[12pt]{article}
 1
\def\as{\alpha_{\rm S}}

\def\n{\!\!}
\def\citenum#1{{\def\@cite##1##2{##1}\cite{#1}}}
\def\citea#1{\@cite{#1}{}}

\def\as{\alpha_{\rm S}}

\def\n{\!\!}
\def\b{\beta}
\def\a{\alpha}
\def\c{\chi   }

\def\D{\Delta}
\def\eps{\epsilon}
\def\g{\gamma}

\def\l{\lambda}
\def\L{\Lambda}
\def\m{\mu}
\def\oa{\omega}

\def\ol{\omega_L}
\def\n{\nu}

\def\p{\phi}
\def\pa{\partial}

\def\ra{\rightarrow}
\def\rh{\rho}
\def\s{\sigma}

\def\ti{\tilde}
\def\v{\vert}

\def\({\left(}
\def\){\right)}

\def\citenum#1{{\def\@cite##1##2{##1}\cite{#1}}}
\def\citea#1{\@cite{#1}{}}

\def\Im#1{\mathop{\rm Im}\{#1\}}
\def\Re#1{\mathop{\rm Re}\{#1\}}

\def\beq{\begin{equation}}
\def\eeq{\end{equation}}
\def\bea{\begin{eqnarray}}
\def\eea{\end{eqnarray}}

\def\eq#1{{eq.~(\ref{#1})}}
\def\eqs#1#2{{eqs.~(\ref{#1})--(\ref{#2})}}
\def\bbbz{{\mathchoice {\hbox{$\sf\textstyle Z\kern-0.4em Z$}}
{\hbox{$\sf\textstyle Z\kern-0.4em Z$}}
{\hbox{$\sf\scriptstyle Z\kern-0.3em Z$}}
{\hbox{$\sf\scriptscriptstyle Z\kern-0.2em Z$}}}}
\def\arnps#1#2#3{  {\it Ann. Rev. Nucl. Part. Sci. }{\bf #1} (19#2) #3}
\def\npb#1#2#3{    {\it Nucl. Phys. }{\bf B#1} (19#2) #3}
\def\plb#1#2#3{    {\it Phys. Lett. }{\bf B#1} (19#2) #3}
\def\prd#1#2#3{    {\it Phys. Rev. }{\bf D#1} (19#2) #3}
\def\prep#1#2#3{   {\it Phys. Rep. }{\bf #1} (19#2) #3}
\def\prl#1#2#3{    {\it Phys. Rev. Lett. }{\bf #1} (19#2) #3}

\def\rmp#1#2#3{    {\it Rev. Mod. Phys. }{\bf #1} (19#2) #3}
\def\zpc#1#2#3{    {\it Z. Phys. }{\bf C#1} (19#2) #3}

\def\sjnp#1#2#3{   {\it Sov. J. Nucl. Phys. }{\bf #1} (19#2) #3}
\def\jetp#1#2#3{   {\it Sov. Phys. }{JETP }{\bf #1} (19#2) #3}

\def\ppjetp#1#2#3{ {\it (Sov. Phys. JETP }{\bf #1} (19#2) #3}
\def\ppjetpl#1#2#3{{\it (JETP Lett. }{\bf #1} (19#2) #3}
\def\zetf#1#2#3{   {\it Zh. ETF }{\bf #1}(19#2) #3}

\def\jpg#1#2#3{        {\it J. Phys}. {\bf G#1}#2#3}
\relax
\begin{document}
\begin{titlepage}
\noindent
 March  1995   \hfill  CBPF-NF-010/95 \\[4ex]\begin{center}
{\Huge \bf  T\,H\,E\,\,\, P\,O\,M\,E\,R\,O\,N:     }   \\[1.4ex]
{\Large \bf YESTERDAY, TODAY and TOMORROW }$^{ \dagger}$  \\[11ex]
\footnotetext{$^{\dagger}$ Lectures given at the III Gleb Watagin School,
Campinas, July 1994 and at the LAFEX,CBPF, Rio de Janeiro, September 1994}

{\large E U G E N E \,\, L E V I N}   $^{*)} $
\footnotetext{$^{*}$ Email: levin@lafex.cbpf.br; levin@fnalv.fnal.gov;
levin@ccsg.tau.ac.il} \\[1.5ex]
{\it  LAFEX, Centro Brasileiro de Pesquisas F\'\i sicas  (CNPq)}\\
{\it Rua Dr. Xavier Sigaud 150, 22290 - 180 Rio de Janeiro, RJ, BRASIL}
\\{\it and}\\
{\it Theory Department, Petersburg Nuclear Physics Institute}\\
{\it 188350, Gatchina, St. Petersburg, RUSSIA}\\[6.5ex]
\end{center}
{\large \bf Abstract:}
These lectures are the review of the main ideas and approaches to
 the structure of the Pomeron. They are divided in three natural parts:

(i) the brief review of the Reggeon Calculus, which was the first attempt to
build the effective theory of the strong interaction at high energy. In spite
of the fact that this approach turns out to be inconsistent and in lectures we
show why, the Reggeon approach was and is the main source of the terminology
 and phenomenology  for  high energy ``soft" interactions.

(ii) the detail description of the QCD approach to high energy interaction.
We try to combine the rigorous approach in perturbative QCD with more simple,
intuitive guess based on general properties of QCD to clarify our expectations
and predictions.

(iii) the outline of my personal opinion what problems will be important in
 the future.

The main reason for the lectures was just the last part to agitate you to
 think about this difficult but because of that interesting problem. The
motto of my lectures is {\it `` there is nothing more exciting than to solve
a difficult problem. Pomeron is the one."}

\end{titlepage}
\section*{ Introduction }
\par
In these lectures I want to give the brief review of the main ideas and
approaches to the structure of the Pomeron. For young generation even the word
``Pomeron"  sounds not familiar. It makes very difficult to explain in the
 introduction  what we are going to discuss. In some sense it is the reason
why we think that we shall start from the short review of the situation with
 the strong interaction in the past because the terminology that we still use
came from so called Regge approach which was in the past the common language
for the phenomenology of high energy strong interaction.

This is why instead of the usual introduction in which I have to explain the
 subject of the talk I am going to provide you some general outline of the main
ideas of the talk.

 I'll start from the review of the Regge approach and
I'll try to give you the basics of this approach, namely the theoretical
 background and the main phenomenological ideas that had been used in
this approach. Just from the beginning I would like to stress that inspite of
the pure phenomenological input the Reggeon approach was and is the main
source  of the ideas even now for the ``soft" high energy interaction. The main
reason for this  is the rich input from the understanding of the general
properties of the scattering amplitude based on the analyticity and symmetry
of strong interaction. The analysis of the theoretical problems made at that
time ( more than 30 years ago) is still fresh and the principle problems
which have been pointed out  are still with us.

After  the introduction in the past and present problems with the ``soft"
 interaction I am going to present what is known about high energy interaction
in perturbative QCD which is the microscopic theory now. We will go back to old
problems and I'll show you what kind of understanding we have reached in
 framework of perturbative QCD (pQCD). This understanding consists of two
 things: the sructure of the Pomeron in perturbative QCD ( so called the BFKL
 Pomeron) and the theory of shadowing correction in the deeply
 inelastic processes.
The main idea of this part of my
 lectures is to show you that we are on the right road but only in
 the beginning. The principle problems are non perturbative ones and we
 have made only the first try to attack them.

The last, third part will be devoted to outline of my personal opinion what
problems will be important in the future. You can easily guess that the
main reason for these lectures was just to agitate you to think about these
difficult but really interesting problems.

As far as concerned the style of these  lectures, we tried to follow the Landau
ten commandments of theoretical physics, at least, the first three of them:

$\bullet$ Only prediction is a theory, only calculation is a result. Don't
believe in a qualitative ``theory".

$\bullet$ Try to solve a problem exactly, if not try to find a small parameter,
if not try again.

$\bullet$ A model is the theory which we apply to the kinematic region where
 we cannot prove that the theory is wrong. Don't be afraid to build a model,
this is the only way to be original.

We would like to make several comments on the quotations. In the first part of
 the  lectures we avoid the references except the most important ones. The
reason
is twofold: first is the fact that everybody can find detail description of
the Reggeon approach in any textbook written in 70's. The second is the idea
 to give you a full information in the body of the lectures about all result
that we are going to take with us in the future. In the second part we give
 sufficiently full list of references.

Concluding this introductory section we want to remind you the motto of these
lectures:
\par
\begin{Large}
{\it There is nothing more exciting than to solve a difficult problem.
Pomeron is the one.}
\end{Large}

\newpage
\begin{Huge}

{\bf YESTERDAY:}
\begin{center}
{\bf REGGEON APPROACH}
\end{center}
\end{Huge}

\section{ The basic ideas.}
\subsection{ Our strategy.}
To understand our theoretical background in the past let me remind you that
 before the year 1974 the study of the high energy asymptotic was the high
 priority job, because we believed that:
\begin{center}
{\it Analyticity\,\,+\,\,Asymptotic\,\,\,=\,\,\,Theory \,\,of \,\,Everything.}
\end{center}
Roughly speaking we needed asymptotic at high energy to specify how many
substructions we have to make in the dispersion relations to calculate the
 scattering amplitude using them.

 Now situation is quite different, we have good microscopic theory ( QCD) and
certainly we have a lot of problems in QCD which  have to be solved. High
 energy asymptotic is only one of many. I think it is time to ask yourselves
why we spend our time and brain trying nevertheless to find the high energy
asymptotic in QCD. My lectures will be an answer to this question, but I would
like to start recalling you the main theorems for the high energy behaviour
of the scattering amplitude which follow directly from the general property of
analyticity and crossing symmetry and should be fulfilled in any microscopic
theory including QCD.
\subsection{ The great theorems.}

{\bf 1.} {\it Optical Theorem}

 \par Our amplitude is
normalized so that
 \begin{equation}
   \frac{d \sigma}{dt} = \pi \vert f(s,t) \vert ^{2}
\end{equation}
Within this normalization the optical theorem says:
 \begin{equation}
   \sigma_{tot} = 4 \pi Im f(s,0)
\end{equation}
Therefore the optical theorem gives us the relationship between the behaviour
of the imaginary part of the scattering amplitude at zero scattering angle and
the total cross section that can be measured experimentally.

{\bf 2.} {\it The Froissart boundary.}

We call the Froissart boundary the following limit of the energy growth of
the total cross section:
\beq
\sigma_{tot} \,\,\,\leq\,\,\,C\,\,ln^2 s
\eeq
where $s$ is the total energy of our elastic reaction: $a(p_a ) + b (p_b)
 \rightarrow a + b$, namely $s \,=\, (p_a \,+\,p_b )^2$.
Coefficient $C$ has been calculated but we do not need to know  its exact
value.
What is really important is the fact that $ C \,\propto \frac{1}{k^2_t}$, where
$k_t$ is the minimal transverse momentum which can be in our reaction. Since
 the minimal mass is the mass of pion in the hadron spectrum the Froissart
theorem says that $ C \,\propto \,\frac{1}{m^2_{\pi}}$. The exact calculation
 gives $C \,= 60 mbn$.

{\bf 3.} {\it The Pomeranchuk Theorem.}

The Pomeranchuk theorem is the manifestation of the { \bf crossing symmetry},
 which
can be formulated as a following statement: one analytic function of two
variables $s$ and $t$ describes the scattering amplitude of
 two different reactions $ a + b \,\rightarrow
a + b$  at $s > 0 $ and $t < 0 $ as well as $ \bar a + b \,\rightarrow \,
\bar a + b $ at $ s < 0  (u = (p_{\bar a} + p_b )^2 > 0 $ and $ t < 0$.

The Pomeranchuk theorem says that the total cross sections of the above two
 reactions should be equal to each other at high energy
if the real part of the amplitude
is smaller than the imaginary its  part.
$$\bullet$$

{\bf Problem 1:} Prove the Pomeranchuk theorem, using the dispersion relation
for the scattering amplitude at $t = 0 $.

\subsection{ Unitarity constraint in impact parameters ($b_t$).}

 The scattering amplitude in $ b_t$-space is defined as
 \begin{equation}
 a(s,b_t) = \frac{1}{2 \pi} \int d^2 q\;\; e^{-i{\bf q\cdot b_t}}
  f(s,t)
\end{equation}
where $ t= - q^{2} $ .
 In this representation
 \begin{equation}
 \sigma_{tot} = 2 \int d^2 b_t \;\; Im a(s,b_t)
\end{equation}
 \begin{equation}
 \sigma_{el} = \int d^2 b_t \;\; \vert a(s,b_t) \vert^{2}
\end{equation}
\par
Using the above notations we can write the $s$- channel unitarity in
 the general form:
\beq
2\,Im\,\,a(s,b_t)\,\,=\,\,\vert a(s,b_t) \vert^2\,\,+\,\,G_{in} (s, b_t)
\eeq
where $ a(s,b_t)$ is elastic amplitude while $G_{in}$ is the contribution
of the all inelastic processes.

It should be stressed that the above constraint has a general solution if
we assume that the scattering amplitude is pure imaginary at high energy,
namely
\begin{equation}
a(s,b_t) = i ( 1 - e^{- \Omega(s,b_t)} )
\end{equation}
where the opacity  $ \; \Omega(s,b) $ is a real function and
has a very simple physical meaning. Indeed,
$$
G_{in} (s, b_t )\,\,=\,\, 1\,\,-\,\,e^{-\,2\, \Omega (s, b_t)}\,\,.
$$
Therefore $e^{ - 2 \Omega}$ is the probability that the incoming particle
 has no inelastic interaction.
$$\bullet$$

{\bf Problem 2:} Prove the Froissart theorem, using the general solution of
the $s$ - channel unitarity taking into account that at large values of $b_t$
the scattering amplitude is small and behaves as $s^N e^{- \mu b_t}$, where
$\mu$ is the mass of the lightest hadron.

\subsection{ The first puzzle.}
The first puzzle can be formulated as a question:{\it What happens with
 resonances with the values of their spin bigger than 1 in exchange?"}
On one hand such resonances have been observed experimentally, on the other
hand the exchange of the resonances with spin $j$ lead to the scattering
 amplitudes which are proportional to $s^j$ where $s$ is the energy of two
colliding hadrons. Such a behaviour contradicts the Froissart boundary.
It means that we have to find the theoretical solution of this problem.

Let me illustrate the problem considering the exchange of
 vector particle ( $j = 1 $) ( see Fig.1).

Introducing Sudakov variable we can expand each vector in the following form:
\beq
q\,\,=\,\,\alpha_q\,p'_1\,\,+\,\,\beta_q\,p'_2\,\,+\,\,q_t
\eeq
where
$$p'^2_1\,=\,p'^2_2\,=\,0$$
and
$$
p_1\,\,=\,\,p'_1\,\,+\,\,\beta\,p'_2\,\,\,\,;\,\,\,\,
p_2\,\,=\,\,\alpha_2\,p'_1\,\,+\,\,\,p'_2\,\,\,\,;
$$
It is easy to find that
$$
\beta_1\,\,=\,\,\frac{m^2_1}{s}\,\,\,\,;\,\,\,\,\alpha_1\,\,=\,\,\frac{m^2_2}{s}
$$
since
$$
p^2_1\,\,=\,\,m^2_1\,\,=\,\,2\,\beta_1 p'_1 \cdot p'_2\,\,=\,\,\beta_1 \,s
$$
{}From equations
$$
p'^2_1\,\,=\,\,M^2_1\,\,\,\,; \,\,\,\,p'^2_2\,\,=\,\,M^2_2
$$
we can find the values of $\alpha_q$ and $\beta_q$. Indeed,
$$
p'^2_1\,\,=\,\,( p_q - q )^2\,=\,(\, ( 1 - \alpha_q) p'_1\,+\,(\beta_1 -
\beta_q)
p'_2\,)^2\,=\,(1 - \alpha_q ) (\beta_1 - \beta_q )s\,-\,q^2_t\,=\,M^2_1
$$
and
$$
p'^2_2\,=\,(1 + \beta_q ) ( \alpha_2 + \alpha_q ) s\,-\,q^2_t\,=\,M^2_2
$$
Therefore
\beq
\alpha_q\,=\,\frac{M^2_2 - m^2_2}{s}\,\,\,;\,\,\,\beta_q\,=\,
\frac{m^2_1 - M^2_1}{s}
\eeq
It is easy to get from the above equations that
\beq
|q^2|\,\,=\,\,|\alpha_q \beta_q s - q^2_t |\,=\,
\frac{( M^2_2 - m^2_2) (M^2_1 - m^2_1 )}{s}\,+\,q^2_t\,\,\rightarrow |_{s
\ra \infty}\,\,q^2_t
\eeq
Now we are prepared to write to expression for the diagram of Fig.1:
\beq
A(s,t)\,\,=\,\,g_1\,g_2 \frac{( p_1 + p'_1 )_{\mu} ( p_2 +p'_2 )_{\mu}}
{ q^2_t + m^2_V}\,=\,g_1\,g_2 \,\frac{4 s}{q^2_t + m^2_V}
\eeq
where $m_V$ is the mass of the vector meson.
\newpage

$$\bullet$$

{\bf Problem 3:} Using the above example show that the exchange of the
resonance
with spin $j$ gives the scattering amplitude equals
$$
A( s, t= - q^2_t )\,=\,g_1\,g_2\,\frac{ ( 4s)^j}{ q^2_t + m^2_R}
$$

$$\bullet$$

{\bf Problem 4:} Show that the exchange of the resonance with spin $j$
leads to the amplitude in $b_t$ which behaves as $ s^j$ $ exp ( - m_R b_t )$ at
large values of $b_t$ and $s$.
\subsection{Reggeons - solutions to the first puzzle.}
The solution to the first puzzle has been found. It turns out that the exchange
of all resonances can be described as an exchange of
the new object - Reggeon and its contribution to the scattering
amplitude is given by the simple function ( see Fig.2):
\beq \label{RGEX}
A_R (s,t)\,\,=\,\,g_1(m_1,M_1, t) \,g_2( m_2,M_2,t) \cdot \frac{ s^{\alpha(t)}
\,\pm \,( - s )^{\alpha (t)}}{ sin \pi \alpha(t)}
\eeq
$\alpha (t)$ is a function of the momentum transfer which we call the Reggeon
 trajectory. The name of the new object as well as the form of the amplitude
came from the analysis of the properties of the scattering amplitude in $t$
 channel using angular momentum representation. However, today it is not
important
the whole history of the approach. What we need to understand and take for
 the future studies are the main properties of the above function which plays
 the crucial role in the theory and phenomenology of the high energy
 interaction.
\section{The main properties of the Reggeon exchange.}
\subsection{Analyticity.}
It is obvious that the Reggeon exchange is the analytic function in $s$, which
has the imaginary part equals to
$$
\pm \,g_1\,g_2 \,s^{\alpha(t)}
$$
in the $s$ - channel and the imaginary part
$$
g_1\,g_2 \,s^{\alpha(t)}
$$
in the $u$ - channel ( for $ \bar a + b \rightarrow \bar a + b$ reaction ).
For different signes in eq.(13) the function has different properties with
 respect to crossing symmetry. For plus ( positive signature) the function is
symmetric while for minus ( negative signature )  it is antisymmetric.
\subsection{ s - channel unitarity}
To satisfy the $s$ - channel unitarity we have to assume that trajectory
 $\alpha (t)  \leq 1 $ in the scattering kinematic region ( $t < 0 $ ).
This is why the exchange of the Reggeons can solve our first puzzle.
\subsection{Resonances.}
Let us consider the same function but in the resonance kinematic region at $t
> 4 m^2_{\pi}$. Here $\alpha (t)$ is a complex function. If $ t \rightarrow
t_0$  $\alpha (t_0) \,=\,j\,=\,
2 k $ where $k = 1,2,3,..$
 the Reggeon
 exchange  has a form
 for the positive signature:
\beq
A_R(s,t)\,\,\rightarrow_{t \rightarrow t_0} g_1\,g_2 \cdot \frac{s^{2k}}{
\alpha'( t_0) \,( t - t_0 ) \,+\,i Im \alpha(t_0)}
\eeq
Since in this kinematic region the amplitude $A_R$ describes the reaction
$ \bar a + a \rightarrow \bar b + b $
$$
s\,\,=\,\,p^2 sin \theta
$$
where $p \,=\,\frac{\sqrt{t_0}}{2}$, the amplitude has a form
\beq
A_R\,\,=\,\,g_1\,g_2\cdot\frac{s^{j}}{\alpha'( t_0) ( t - t_0 ) \,+
\,i Im \alpha( t_0 ) }\,\,=\,\, \frac{g_1 g_2 }{\alpha' (t_0)}
\cdot \frac{p^{2j} sin^j \theta}{t - t_0 + i \Gamma}
\eeq
where the resonance width $\Gamma = \frac{Im \alpha (t_0)}{ \alpha' (t_0)}$.

Therefore the Reggeon gives the Breit - Wigner  amplitude of the resonance
 contribution
at $t > 4 m^2_{\pi}$. It is easy to show that the Reggeon exchange with the
negative signature describes the contribution  of a resonances with odd
spin $j = 2k + 1 $.
\subsection{Trajectories.}
We can rephrase the previous observation in different words saying that
the Reggeon describes the family of resonances that lies on the same trajectory
$\alpha(t)$. It gives us a new approach to the classification of the
resonances,
which is quite different from usual $SU_3$ classification. Fig.3 shows the
bosonic resonances classified according the Reggeon trajectories.
The surprising experimental fact is that all trajectories are the straight
 lines
\beq
\alpha( t ) \,\,=\,\,\alpha (0) \,\,+\,\,\alpha'\,t
\eeq
with the same value of the slope $\alpha' \simeq 1 GeV^{- 2}$.

We would like to draw your attention to the fact that this simple linear form
comes from two experimental facts: 1) the width of resonances are much smaller
than their mass ($ \Gamma_i \,\ll \, M_{R_i}$) and 2) the slope of the
trajectories which is responsible for the shrinkage of the diffraction peak
turns out to be the same from the experiments in the scattering kinematic
 region.

 The $SU_3$
means in terms of trajectories that
$$\alpha_{\rho} (0)\,=\,
\alpha_{\omega} (0)\,=\,
\alpha_{\phi} (0)$$
in addition to the same value of the slope. The simple estimates show that the
 value of the intercept $\alpha(0) \,\simeq 0.5$ and therefore the exchange of
the Reggeons give the cross section falling down as function of the energy
without any violation of the Froissart theorem.

\subsection{ Definite phase.}
The Reggeon amplitude of \eq{RGEX} can be rewritten in the form:
\beq
A_R\,\,=\,\,g_1 g_2 \eta_{\pm}  s^{\alpha(t)}
\eeq
where $\eta$ is the signature factor
$$
\eta_{+}\,\,=\,\,ctg\frac{\pi \alpha(t)}{2} \,\,+\,\,i
$$
$$
\eta_{-}\,\,=\,\,tg \frac{\pi \alpha(t)}{2}\,\,-\,\,i
$$
It means that the exchange of the Reggeon brings very definite phase
 of the scattering amplitude. This fact is very important especially for
 the description of the interaction with a polarized target.
\subsection{Factorization.}
The amplitude of \eq{RGEX} has a simple factorized form in which all
 dependences on the particular properties of colliding hadrons are concentrated
in the vertex functions $g_1$ and $g_2$. To make it clear let us rewrite this
factorization property in an explicit way:
\beq
A_R\,\,=\,\,g_1 (m_1,M_1,q^2_t) \,g_2 (m_2,M_2,q^2_t) \cdot \eta_{\pm} \cdot
s^{\alpha( q^2_t)}
\eeq
For example this form of the Reggeon exchange means that if we try to
 describe the electron  deep inelastic scattering with the target
 through the Reggeon
 exchange, only the vertex function should depend on the value of the
 virtuality of photon ($Q^2$) while the energy dependance does not depends on
$Q^2$.

$$\bullet$$
 {\bf Problem 5:} Show that the Reggeon exchange has the following form in
$b_t$:
$$
A_R(s,b_t)\,\,=\,\,g_1(0)\,g_2(0) s^{\alpha(0)} \cdot \frac{1}{4\pi
 ( R^2_1 + R^2_2 n+ \alpha' \ln s} \cdot e^{- \,\frac{b^2_t}{4( R^2_1
+ R^2_2 + \alpha' \ln s)}}
$$ if we assume the simple exponential parameterization for the vertices:
$$
g_1 (q^2_t) = g_1(0) \,e^{- R^2_1 \,q^2_t}
$$
$$
g_2(q^2_t) = g_2(0) \,e^{- R^2_2 \,q^2_t}
$$
\subsection{Shrinkage of the diffraction peak.}
Using the linear trajectory for the Reggeons it is easy to see that the
elastic cross section due to the exchange of the Reggeon can be written
 in the form:
\beq
\frac{d \sigma_{el}}{d t}\,\,=\,\,g^2_1(q^2_t)\cdot q^2_2( q^2_t)
\cdot s^{2( \alpha(0) - 1)} \cdot e^{- 2 \alpha'(0)\,\ln s\,q^2_t}
\eeq
The last exponent reflects the phenomena that is  called as the shrinkage of
 the diffraction peak. Indeed, at very high energy  the elastic cross section
is concentrated at values of $q^2_t < \frac{1}{\alpha' \ln s}$. It means that
the diffraction peak becomes narrower at higher  energies.
\section{Analyticity + Reggeons.}
Now we can come back to the main idea of the approach and try to construct the
amplitude from the analytic properties and the Reggeon asymptotic at high
 energy. Veneziano \cite{VEN} suggested the amplitude that satisfies
 the following constraint: it is
the sum of all resonances in $s$ - channel with zero width and simultaneously
the same amplitude is the sum of the exchanges of all possible Reggeons.
Taking the simplest case of the scattering of scalar particle, the Veneziano
 amplitude looks as follows:
\beq
A\,\,=\,\,g\,[\,V(s,t)\,+\,V(u,t)\,+\,V(s,u)\,]
\eeq
where
\beq
V(s,t)\,\,=\,\,\frac{\Gamma( 1 - \alpha(t) ) \Gamma( 1 - \alpha (s ) )}{\Gamma
( 1 - \alpha(t) - \alpha(s) )}
\eeq
One can see that the Veneziano amplitude has resonances at $\alpha(s) = n + 1$
where $n = 1,2,3...$, since $\Gamma (z) \rightarrow \frac{1}{z + n}$ at
 $z \rightarrow - n $. At the same time
$$
A\,\,\rightarrow_{s \rightarrow \infty}\,\,\Gamma( 1 - \alpha(t) ) [ (
- \alpha(s))^{\alpha(t)} \,\,+\,\, (- \alpha(u) )^{\alpha(t)} ]
$$
and therefore reproduces the Reggeon exchange at high energies.

This simple model was the triumph of the whole approach
 showing us how we can construct the theory using analyticity and asymptotic.
The idea was to use the Veneziano model as the first approximation or in other
word as a new Born term in the theory and to try to build the new theory
 starting with the new Born Approximation. The coupling constant $g$ turns out
to be dimensionless and smaller than unity. This fact certainly also encouraged
the theoreticians in 70's to try this new approach.
\section{The second and the third puzzles: the Pomeron?!}
The experiment shows that:

{\bf 1.} {\it There is no particles ( resonances) on the Reggeon trajectory
with
the value of the intercept which is close to unity ( $\alpha(0) \rightarrow
 1$).} As it has been mentioned the typical highest value of the intercept
is $\alpha(0) \sim 0.5$ which generates the cross section of the order of
$\sigma_{tot} \propto s^{\alpha(0) - 1} \propto s^{-\frac{1}{2}}$.

{\bf 2.}{\it The total cross section is approximately constant at high energy.}
It means that we have to assume in the framework of the Reggeon approach
that there is the Reggeon with the intercept close to 1.

The above two statement we call the second and the third puzzles that have to
 be solved in theory. The strategy of the approach was to assume that the
Reggeon with the intercept close to 1 exists and tried to understand how
 and why it would be different from other Reggeons, in particular, why there
 is no particle on this trajectory.
Now, let me introduce for the first time the word Pomeron. The first
 definition of the Pomeron:

{\bf The Pomeron is the Reggeon with $\alpha(0) - 1 \,\equiv \,\Delta\,\ll
\,1$}

The name of Pomeron was introduced after russian physicist  Pomeranchuk who
did a lot to understand this miracle object. By the way,  the general name of
the Reggeon was given after the italian physicist Regge, who gave a beautiful
theoretical arguments why such objects can exist in the quantum mechanics and
the  field theory \cite{REGGE}.

Let me summarize what we know about the Pomeron from the experiment
( see for example ref. \cite{DL}):

1. $\Delta \,\simeq\,0.08$

2.$\alpha'(0)\,\simeq\,0.25 GeV^{-2} $

Donnachie and Landshoff gave an elegant description of almost all existing
experimental data using the hypothesis of the Pomeron with the above parameters
of its trajectory. However, we have to find the theoretical approach how to
describe the Pomeron and why it is so different from other Reggeons.
\section{The Pomeron structure in the parton model.}
\subsection{The Pomeron in the Veneziano model.}
As has been mentioned the Pomeron does not appear in the new Born term of our
approach. Therefore the first natural idea was to make an attempt to calculate
the next to Born approximation in the Veneziano model to see can the Pomeron
appear in it. The basic equation that we want to use is graphically pictured
 in Fig.5, which is nothing more than the optical theorem. However we have to
 know the Born approximation for the amplitude of production of $n$ particle.

However to understand the main properties and problems which can arise in this
approach let us calculate the contribution in equation of Fig. 5 of the first
 two particle state. This contribution is equal to:
\beq
\sigma_{tot}\,\,=\,\,s^{2(\alpha_R(0) - 1 )} \,\,\int d^2 q'_t \Gamma^2
(1-\alpha(q'^2_t)) \cdot e^{ - 2\alpha' q'^2_t \ln s}
\eeq
since
$$
\Gamma( 1 - \alpha( q'^2_t))\,\propto\,
(\frac{\alpha(q'^2_t)}{e})^{2 \alpha(q'^2_t)}
$$
one can see that the essential values of $q'$ in the integral is rather big,
 of the order of $q'^2_t \,\sim s $. It means that we have to believe in the
Veneziano amplitude at the large value of momentum transfer. Thus the lesson
that we have learned from this exercise is the following:

{\it To understand the Pomeron structure we have to understand better the
structure of the scattering amplitude at large values of the momentum
 transfer or in other words we should know the interaction at small distances.}

\subsection{The parton model.}
Feynman \cite{FE} and Gribov \cite{GR} suggested the simple model for
 the scattering amplitude at small distances that manifests itself in the
deep inelastic scattering \cite{BJ},so called the parton model.
 Let us assume that in the production
 amplitude the mean transverse momentum of the secondary particles does
 not depend on the energy ( $ k_{it} = Const (s) $ ( see Fig.6). If it is so,
the main contribution to the equation of Fig.5 comes from very specific
region of integration. Indeed, the total cross section can be written as
 follows:
\beq
\sigma_{tot}\,\,=\,\,\Sigma_n \int |M^2_n ( x_i p, k_{ti})|
 \Pi_i \frac{ d x_i}{x_i}\,\, d^2 k_{ti}
\eeq
where $x_i$ is the fraction of energy that carried by the $i$-th particle.
Let's call all secondary particle partons.

It is clear that the biggest contribution in the above equation comes from
 the region of integration with strong ordering in $x_i$ for all produced
 partons,namely
\beq
x_1\,\gg \,x_2\,\gg .....\gg \,x_i\,\gg\, \,x_{i + 1} \,\gg .....\gg \,x_n\,=\,
\frac{m^2}{s}
\eeq
Integrating in this kinematic region we can put all $x_i= 0 $ in the amplitude
$M_n$. Finally,
\beq
\sigma_{tot}\,\,=\,\,\Sigma_n \int \Pi_i d^2 k_{ti} |M^2_n (k_{ti})| \,
\int^1_{\frac{m^2}{s}} \,\frac{d x_1}{x_i} .....\int^{x_{i -
1}}_{\frac{m^2}{s}}
\,\frac{d x_i}{ x_i}\,...\int^{x_{n - 1}}_{\frac{m^2}{s}} \,\frac{d x_n}{ x_n}
\eeq
$$=\,\,\Sigma_n \int \Pi_i d^2 k_{ti} |M^2_n (k_{ti})| \,\cdot\,
\frac{1}{n!}\,\ln^n s
$$
This equation shows one very general property of the high energy interactions,
 namely the longitudinal coordinates ( $x_i$) and the transverse ones
($k_{ti}$)
turns out to be separated and should be treated differently. In some sense the
above equation reduced the problem of high energy behaviour of the total
cross section to the calculation of the amplitude $M_n$ which depend only on
transverse coordinates. Assuming, for example, that $\int \Pi d^2 k_{ti}
|M^2_n( k_{ti})|^2\,\propto \frac{1}{m^2} g^n$ we can derive from the eq.(25)
\beq
\sigma_{tot}\,=\,\frac{1}{m^2}\,\Sigma_n\,\frac{g^n}{n!} \,\ln^n s\,=\,
\frac{1}{m^2}\cdot s^{g}
\eeq
which looks just as Pomeron - like behaviour.

$$\bullet$$
{\bf Problem 6:} Show that in $g \phi^3$ - theory where $\phi$ is the scalar
particle with mass $m$ the total cross section is equal to
$$
\sigma_{tot}\,\,=\,\,\frac{g^4}{m^2 4\pi s} \,\,s^{\frac{g^2}{4\pi m^2}}
$$

\subsection{Random walk in $ b_t$.}
The simple parton picture reproduces also the shrinkage of the diffraction
peak.
Indeed, due to the uncertainty principle
\beq
\Delta b_{ti} \,\,k_{ti}\,\sim \,1
\eeq
or in different form
$$
\Delta b_{ti}\,\sim \frac{1}{ < k_t >}
$$
Therefore after each emission the position of the parton will be shifted on the
value of $\Delta b_t$ which is the same in average. After $n$ emission we
 have the picture given in Fig.7, namely the total shift in $b_t$ is equal to
\beq
b^2_{tn}\,\,=\,\,\frac{1}{< k_t>^2} \cdot n
\eeq
which is typical answer for the random walk in two dimensions ( see Fig.7).
The value of the average number of emission $n$ can be estimated from the
expression for the total cross section, because
$$
\sigma_{tot}\,=\,\frac{1}{m^2}\,\Sigma_n\,\frac{g^n}{n!} \,\ln^n s\,=\,
\frac{1}{m^2}\cdot s^{g}\,\,\sigma_0\,\Sigma_n \frac{ <n>^n}{n!}
$$
and the value of $< n> \simeq g \ln s$.
If we substitute this value of $<n>$ in the eq.(0) We get the radius of
 interaction
$$
R^2\,\,=\,\,<b^2_{tn}>\,\,\frac{g}{<k_t>^2} \cdot \ln s\,\,
=\,\,\alpha'\,\ln s
$$
\section{Reggeon Calculus.}
\subsection{The main idea of the approach.}
The Reggeon Calculus was the first attempt to build the effective theory at
 high energy with the goal to define the asymptotic behaviour of the
scattering amplitude. The brick from which we wanted to do this was the
Pomeron.
However, it turns out that the simple hypothesis that the Pomeron gives you
 the asymptotic behaviour of the amplitude  is not correct. We have to take
 into account the interaction of the Pomerons. To illustrate this fact let us
consider so called the triple Pomeron interaction which related to the process
of the diffraction dissociation ( see Fig.8 ).

The cross section of the diffraction dissociation ( single diffraction SD)
 can be written in the form:
\beq
\frac{M^2 d \sigma_{SD}}{d M^2}\,\,=\,\,\frac{\sigma_0}{2\pi R^2_1
 (\frac{s}{M^2})}\,\cdot\,( \frac{s}{M^2})^{2 \Delta} \,\cdot\,\gamma\,\cdot\,
(\frac{M^2}{s_0})^{\Delta}
\eeq
where $\sigma_0 $ is the total cross section at $s=s_0$ due to exchange of one
Pomeron and $R^2_1 = 2 B_{SD}$, where $B_{SD}$ denotes the slope of the SD
cross section. It should  be stressed that at very high energy when
$\alpha'_{P} \,\gg \,R^2_0 $,  $R^2_1 \,\rightarrow \, 4\alpha'_{P}\,\ln
(s/M^2)$

Integrating over $M^2$ the SD cross section one can see that the total
cross section of SD
$$
\sigma_{SD}\,\,\propto\,\, (\frac{s}{s_0})^{2\Delta}\,\,\gg\,\,\sigma_{tot}\,
\propto\, ( \frac{s}{s_0})^{\Delta}\,\,.
$$
It means that we have to consider the interaction of the Pomerons to get the
correct asymptotic at high energy.
$$\bullet$$

{\bf Problem 7:} Show that energy $s'$ in Fig.8 is proportional
 to  $\frac{s}{M^2}$.

$$\bullet$$
{\bf Problem 8:} Show that the total cross section of SD is larger that
one Pomeron exchange even at $\Delta = 0 $.

\subsection{Reggeon Diagram Technique.}
It turns out that the simplest way to deal with the Pomeron interaction is
to use the Mellin transform of the amplitude \cite{GR1}
\beq
A (\omega, t)\,\,=\,\,\int^{\infty}_0 \,\,s^{- 1 - \omega} d\,s \,\,Im A(s, t )
\eeq
\par
$$\bullet$$

{\bf Problem 9:} Show that the Mellin transform of the single Pomeron exchange
gives
$$A_P \,\,=\,\,\frac{1}{\omega\,-\,\Delta\,-\,\alpha'_P \,q^2_t} $$
where $t = - q^2_t$.

$$\bullet$$

Using the above Pomeron propagator it is easy to write  any diagram for
 Pomeron interaction. For example, for the triple Pomeron interaction the
diagram of Fig.9 can be written in the form:
\beq
A_{3P}\,\,=\,\,g\,\frac{1}{\omega - \Delta - \alpha'_P q^2_t}\,\gamma\,
\frac{1}{ \omega - 2 \Delta - \alpha'_P \,[ \,(q - k )^2_t \,+\,k^2_t\,]}\,N
\eeq
\subsection{The effective Lagrangian.}
The above examples show that the new effective theory can be build as the
 theory of the interaction of the effective degrees of freedom (Pomerons) in
 (1 + 2) dimensions. However, it is obvious that we cannot built the closed
 theory without specification what kind of interaction we want to take into
account. On this pure phenomenological stage we have no selection rules and
at first sight have to include any possible interactions. The idea was to
 start from triple Pomeron interaction, to solve the problem and to check back
whether the more complicated interactions would be essential.

For triple Pomeron interaction the effective Lagrangian looks very attractive:
\beq
{\it L}\,\,=\,\,-\,\frac{1}{2}\,[\,\phi^{+} \frac{\partial \phi}{\partial t}\,
+\,\frac{\partial \phi^{+}}{\partial t} \phi\,]\,+\,\Delta \phi^{+}\,\phi
\,-\,\frac{1}{2}\,\gamma\,[\,\phi^{+}\,+\,\phi\,]\phi^{+} \phi
\eeq
where $t = ln s $ and the propagator of the Pomeron is defined as the Green
function of the Lagrangian without interaction ($ \gamma$ = 0 ).
We will discuss a bit later the sign minus in front of the last term.

The attempts to solve the effective theory were very interesting and
 we learned a lot, but we will not discuss them because we will show later
that the whole approach is deadly sick.

\subsection{The AGK cutting rules.}
The AGK cutting rules \cite{AGK} establish the generalization of the optical
theorem on the case of the multi Pomeron interaction and in the simplest
 example of two Pomeron exchange they are pictured in Fig.11.
The meaning of the AGK cutting rules is very simple. Each Pomeron interaction
contribution to the total cross section really corresponds to the cross
 section of the different processes. In the case of the two Pomeron exchange
these processes are (i) the diffraction dissociation ( $\sigma^{(0)}$ )
, (ii) the production
of the secondary hadrons with the same multiplicity as for one Pomeron exchange
( $ \sigma^{(1)}$ )
and (iii)  the production of the secondary hadrons with multiplicity in two
times
 larger than for one Pomeron exchange ( $\sigma^{(2)}$ ).

The AGK cutting rules claim:
\beq
\sigma^{(0)}\,\div\,\sigma^{(1)}\,\div\,\sigma^{(2)}\,\,=\,1\,\div\, - 4 \,
\div\,2
\eeq
Two important consequences follow from the AGK cutting rules:

1. The total cross section of the diffraction dissociation (DD) is equal to
 the contribution of two Pomeron exchange to the total cross section
( $\sigma^{(2P)}_{tot}$ ) with opposite sign:
$$
\sigma^{(DD)}\,\,=\,\,-\,\sigma^{(2P)}
$$

2. The two Pomeron exchange does not contribute to the total inclusive cross
 section in central kinematical region. Indeed only two processes $
\sigma^{(1)}$ and $\sigma^{(2)}$ are the sources of the produced particle in
 the central region. It means that the total inclusive cross section due
to two Pomeron exchange is equal to
$$
\sigma_{inc}\,=\,\sigma^{(1)}\,+\,2\,\sigma^{(2)}\,=\,0
$$
Factor 2 comes from the fact that the particle can be produced from two
 different parton showers (see Fig.11).

Now let me give you a brief proof of the AGK cutting rules for the case of
hadron - deuteron interaction. For simplicity let us assume that $G_{in}(b_t)
\,=\,\kappa \,=\, Const (b_t)$ for $b_t < R_N$ and $G_{in} = 0 $ for $b_t >
R_N$
, so the inelastic cross section for hadron -
 nucleon interaction is equal to
$$
\sigma^{inel}_N\,=\,\kappa \,S
$$ where S is the area of the nucleon  ( S = $\pi R^2_N$ ).
To calculate the elastic cross section we need to use the unitarity constraint
in $b_t$ ( see eq.(7)) and the answer is
$$
\sigma^{el}_N \,=\,( \frac{\kappa}{2})^2 S
$$
Note that the flux of incoming particles after the first interaction becomes
$ ( 1 - \kappa ) $ we can calculate the total inelastic interaction with the
 deutron, namely
$$
\sigma^{inel}_D\,\,= \kappa S \,\,+\,\, \kappa ( 1 - \kappa ) S\,\,=\,\,
 2 \,\sigma^{inel}_N - \kappa^2 S
$$
Using unitarity we have to calculate the elastic cross section for the
 interaction with deuteron which is equal to
$$
\sigma^{el}_D\,\,=\,\, (\frac{2 \kappa}{2})^2 S
$$
Therefore the total cross section for hadron - deutron interaction can be
presented in the form:
$$
\sigma^{tot}_D\,\,= 2 \,\sigma^{tot}_N \,- \,\Delta \sigma
$$
where
$$
\Delta \sigma\,\,=\,\,\Delta \sigma^{inel} \,+\,\Delta \sigma^{el}\,=
\,-\,\frac{\kappa^2}{2}\,S\,=\,-\,\frac{ (\sigma^{inel}_N )^2}{2 \pi R^2}
$$
Everybody can recognize the usual Glauber formula for hadron - deutron
interaction.

As has been mentioned we have two sources of the inelastic cross section  which
are pictured in Fig. 13 for the case of the deutron. The cross section of the
inelastic process with double multiplicity is easy to calculate, because it is
equal to the probability of two inelastic interaction:
$$
\Delta \sigma^{(2)}_D\,\,=\,\,\kappa^2 \cdot S
$$
To calculate;ate $\sigma^{(1)}_D$ we have to remember that
$$
\Delta \sigma^{inel}\,\,=\,\,\sigma^{(2)}\,\,+\,\,\sigma^{(1)}\,\,=\,\,-
\kappa S
$$
Therefore $\sigma^{(1)} \,=\,- 2 \kappa S$. Remembering that $ \Delta
\sigma^{(0)}\,=\,\sigma^{el}_D \,-\,2 \sigma^{el}_N$
 we get
\begin{Large}

\begin{tabular}{|l|l|l|r|}
   \hline
 n  &  0
&   $<n>_N $     &   2$<n>_N$  \\   \hline
$\Delta \sigma_{tot}$  & $ \frac{\kappa^2}{2} S $ & $ - 2 \kappa^2 S $
& $\kappa^2 S$ \\ \hline
\end{tabular}
\end{Large}
\par
Therefore we get the AGK cutting rules for the hadron - deuteron interaction,
since  $\sigma^{tot}_D \,=\,2 \sigma^{tot}_N$ corresponds to one Pomeron
exchange  and the correction to this simple formula just originated from the
two
Pomeron exchange in our Reggeon approach. However the above discussion, I hope,
shows you that the AGK cutting rules have more general background than the
Reggeon approach. For example, they hold in QCD providing the so called
factorization theorem.

Let me give here the general formula for the AGK cutting rules \cite{AGK}.
If we have the contribution to the total cross section with the exchange of
$\nu$ Pomerons ( $\sigma^{\nu}_{tot}$),  the cross sections of  the process
with the multiplicity of produced particles equal $\mu <n>$
  which are generated by the
 same diagram (see Fig.13) are equal\footnote{
  $<n>$ is the
average multiplicity in one Pomeron exchange}:

\beq
 \frac{\sigma^{(\mu)}}{\sigma^{\nu}_{tot}}\,|_{\mu \neq	 0}\,\,=\,\,
( - 1 )^{\nu - \mu} \cdot \frac{ \nu!}{\mu! \,( \nu  - \mu)!} \cdot 2^{\nu}
\eeq
while for $\mu = 0 $ the ratio is equal
\beq
 \frac{\sigma^{(0)}}{\sigma^{\nu}_{tot}}\,\,=\,\,
( - 1 )^{\nu} \cdot [\, 2^{\nu - 1}\,-\,1\,]
\eeq
For the first 5 exchange you can find these factors in the following table.

\begin{tabular}{|l|r|r|r|r|r|r|}
   \hline
 $\nu \setminus \mu$  &  0 & 1 & 2 & 3 & 4& 5 \\   \hline
 1 & 0 & 1 & 0 & 0 & 0 & 0  \\ \hline
2 & 1 & - 4 & 2 & 0 & 0 & 0 \\ \hline
3 & - 3 & 12 & - 12 & 4 & 0  & 0 \\ \hline
4 & 7 & - 32 & 48 & - 32 & 8 & 0 \\ \hline
5 & - 15 & 80 & - 160 & 160 & - 80 & 16 \\ \hline
\end{tabular}
\subsection{ Different processes in the Reggeon Approach.}
The AGK cutting rules together with Mueller theorem \cite{MU} establish
 the relationship between the contribution of many Pomeron exchanges and
different exclusive and inclusive processes in the Reggeon Approach. Here
we want to write down several examples of different processes that can be
treated on the same footing in the reggeon approach.
\subsubsection{Total cross section.}
As has been mentioned in one Pomeron exchange approximation the total cross
section  is given by the following expression ( see Fig. 15 \footnote{In
Fig. 15 we use a little bit different notation, namely $\beta (t )$ instead
of g (t ) and g ( t) for $G_{3P}( t )$. We did this because we had nice
 prepared pictures in such notations. Hope that this fact will not lead to
misunderstanding.}):
\beq
\sigma_{tot}\,\,=\,\,4 \pi g_1(0) \,g_2(0)  (\frac{s}{s_0} )^{\Delta}\,\,
\sigma_0 ( \frac{s}{s_0})^{\Delta}
\eeq
We would like to remind you that  multi  Pomeron exchange is very essential
 in the total cross section but we postpone the discussion of their
 contribution to the second part of our lectures.
\subsubsection{Elastic cross section.}

$$\bullet$$
{\bf Problem 10:} Show that for one Pomeron exchange the total elastic cross
 section is equal to
$$
\sigma_{el}\,\,=\,\,\frac{\sigma_{tot}}{16 \pi B_{el}}
$$ where
 $$B_{el} \,\,=\,\,2 R^2_{01}\,+\,2 R^2_{02} \,+\, 2 \alpha'_P \ln s/s_0$$
in the exponential parameterization for the vertices $g(t)$ ( see Problem
5).
\subsubsection{Single diffraction dissociation}
The cross section of the single diffraction (see Fig.15) has the following
form:
\beq
\frac{M^2 d \sigma_{SD}}{d M^2}\,\,=\,\,\frac{\sigma_0}{2\pi R^2_1
 (\frac{s}{M^2})}\,\cdot\,( \frac{s}{M^2})^{2 \Delta} \,\cdot\,[\,G_{3P} (0)
\,\cdot\,
(\frac{M^2}{s_0})^{\Delta} \,+\,G_{PPR}(0)
 (\frac{M^2}{s_0} )^{ \alpha_R (0) - 1}\,]
\eeq
\par
$$\bullet$$

{\bf Problem 11:} Show that in exponential parameterization for vertices
 (see Problem 5)
$$
R^2 (\frac{s}{M^2})\,\,=\,\,2 R^2_{01} \,+\,r^2_0  \,+\,4 \alpha'_P \ln (s/M^2)
$$
where $r^2_0$ is the radius of the triple Pomeron vertex which experimentally
is very small ($ r^2_0 \leq 1 GeV^{-2} $).

$$\bullet$$

One can see that we include in the expression for the cross section of
 the single diffraction also the contribution of the secondary Reggeon
trajectory which describes the behaviour of the cross section at rather
 small values of produced mass ($M^2$).

\subsubsection{ Double diffraction dissociation.}
$$\bullet$$

{\bf Problem 12:} Show that the cross section of the double diffraction
 dissociation ( DD ) process (see Fig.15) is equal to:
\beq
\frac{M^2_1 M^2_2 d^2 \sigma_{DD}}{d M^2_1 d M^2_2}\,\,=\,\,
\frac{\sigma_0}{ 2 \pi R^2_0 (\frac{s s_0}{M^2_1 M^2_2} )} \cdot
G^2_{3P} (0) \cdot ( \frac{s s_0}{M^2_1 M^2_2} )^{2 \Delta} \cdot
( \frac{M^2_1}{s_0})^{\Delta}
  \cdot
( \frac{M^2_2}{s_0})^{\Delta}
\eeq
in the region of large values of produced masses ( $M_1$ and $M_2$ ).

$$\bullet$$

{\bf Problem 13:} Show that $R^2_0$ in the above equation is equal
$$
R^2_0\,\,=\,\,\frac{1}{2} r^2_0 \,\,+\,\, 4 \alpha'_P \ln ( s s_0/M^2_1 M^2_2)
$$
in exponential parameterization of the vertices ( see Problem 5 ).

$$\bullet$$

Comparing the cross section for double and  single diffraction as well as
for elastic cross section one can get the following factorization relation:

\beq
\frac{M^2_1 M^2_2 d^2 \sigma_{DD}}{d M^2_1 d M^2_2}\,\,=\,\,
\frac{\frac{M^2_1 d \sigma_{SD}}{d M^2_1} \frac{M^2_2 d \sigma_{SD} }{d M^2_2}}
{ \sigma_{el}}\,\cdot\,\frac{R^2_1 (\frac{s}{M^2_1}) R^2_2 (\frac{s}{M^2_2})}
{R^2_0 (\frac{s s_0}{M^2_1 M^2_2}) R^2_{el} (s)}
\eeq
where
$$R^2_{el} = 2 R^2_{01} + 2 R^2_{02} + 4 \alpha'_p \ln (s/s_0) $$
The structure of the event which corresponds to double diffraction is pictured
in Fig.15 in lego -  plot. In the region of rapidity $\Delta y $ we have
 no produced particles.
\subsubsection{ Central Diffraction.}
In this process we produced the bunch of secondary particle in the central
 rapidity region, while there are no secondary particles in other regions in
rapidity.

 The first diagram that describes this process gives the answer:
\begin{eqnarray}
   M^{2} \frac{d \sigma}{d M^{2}} =
        \frac{1}{2 \alpha^{\prime}} \sigma_{PP}(M^{2})g^{4}_{0}
         (\frac{s}{M^{2}})^{2 \Delta}
 e^{\frac{1}{2}(R^{2}_{0} + \alpha^{\prime} ln\frac{s}{M^{2}})q^{2}}
 \frac{1}{R^{2}_{0} + \alpha^{\prime}ln\frac{s}{M^{2}}} \nonumber \\
\cdot ln [ \frac{R^{2}_{0} + 2 \alpha^{\prime} ln\frac{s}{M^{2}}}
    {R^{2}_{0}}]
\end{eqnarray}
The above examples demonstrate how we can calculate the exclusive processes
using several phenomenological inputs in the Reggeon Approach, however
all these processes are crucially affected by multi Pomeron contributions
and this subject we are going to discuss later.
\subsubsection{Inclusive cross section.}
As has been discussed the inclusive production does not depend on multi
 Pomeron exchanges. Using Mueller technique (see Fig.16) which is a
 generalization of the optical theorem for more complicated than total
cross sections cases  the inclusive cross section can be written
 in a very simple form:
\beq
\frac{d \sigma}{d y_c} \,\,=\,\, a \cdot \sigma_{tot}
\eeq
where $a$ is the new vertex for the emission of the particle $c$.
The inclusive reaction that is under consideration is
$$
a\,\,+\,\,b\,\,\rightarrow\,\,c(y)\,\,+\,\,anything
$$
\subsubsection{Two particle rapidity correlation.}
The Reggeon approach we can use for the estimates of the two particle
 rapidity correlation function, which is defined as:
\beq
R \,\,=\,\,\frac{\frac{d^2 \sigma( y_1, y_2 )}{\sigma_{tot} d y_1 d y_2}}{
\frac{d \sigma(y_1)}{ d y_1} \frac{d \sigma (y_2)}{ d y_2}}\,\,-1
\eeq
where  $ \frac{d^2 \sigma}{d y_1 dy_2}$ is the double inclusive
cross section of the reaction:
$$
a\,\,+\,\,b\,\,\rightarrow \,\,1 (y_1)\,\,+\,\,2 (y_2)\,\,+ \,\,anything
$$
\par
$$\bullet$$

{\bf Problem 14:} Show using the AGK cutting rules
that the correlation function R is equal:
\beq
R (\Delta y = |y_1 - y_2|)\,\,=\,\,\frac{a^2_{PR}}{a^2_{PP}}\,\cdot\,
e^{ ( 1 - \alpha_R (0) ) \,\Delta y }\,\,+\,\,2\,\frac{\sigma^{(2P)}_{tot}}
{\sigma^{(P)}_{tot}}
\eeq
All notations are clear from Fig.16.
$$\bullet$$
It should be stressed that the Reggeon approach gives the estimates for
 the correlation length ( $ L{cor}$) and the the strength of the long range
correlations ( note, that the second term does not depends on rapitidies
 of produced hadrons). Namely, we can rewrite the above formula in the form:
$$
R \,\,=\,\, SR \cdot e^{- \frac{\Delta y}{L_{cor}}}\,\,+\,\,LR
$$
and from our Reggeon formula we have $L_{cor} \simeq 2 $. To evaluate the
  second term we have to develop some model for including the multi Pomeron
exchanges in the calculation of the total cross section. We will do this
in the second part of the lectures.

\subsection{Hopes.}
The main idea of the Reggeon Calculus was to build the effective theory
for the strong interaction at high energy starting from the effective
 Lagrangian and the AGK cutting rules. The last one played a very crucial role
giving us the possibility to look insight of the inelastic interaction using
only general ideas on the structure of the Pomeron.

Of course one defect of the approach has been seen from the beginning, namely
the absence of a theoretical idea how to select the interaction between
 Pomerons.  However we could hope that the future theory will provide us
such selection rules and we will be able to adjust the developed formalism.

\section{ The death of the Reggeon Approach.}
In 1974 - 1975 in ref. \cite{TTW} was shown that all our hopes were in vain.
T.T. Wu and McCoy and S.Matinian and A. Sedrokian proved that in the general
 class of the theories ( all theoretical model that existed at that time)
our hope that the Pomeron and Pomeron interaction can describe the high
 energy asymptotic is not correct.

 Let me illustrate this point using the
parton model. In this model the Pomeron is the sum of ``ladder" diagrams
while the two Pomeron exchange corresponds to the diagram of Fig.17 where
the partons are produced and absorbed by the same ``ladders". These
diagrams have the probabilistic interpretation and look very natural in the
 parton model. However it turns out that the diagrams where the secondary
 partons are produced by one ``ladder" and are absorbed by the second one
give bigger contribution. These diagrams have not been included neither in one
Pomeron exchange nor in two Pomeron exchange.

It means that our hope to reduce the whole theory to Pomeron interaction
can not be right or at least we have to build more complicated picture for the
Pomeron structure, which is impossible to do without detailed microscopic
 theory.

It means also that the AGK cutting rules we cannot apply for the  process
of the Pomeron interaction. For example, if we want to calculate
 the inclusive production of the hadron in the single diffraction process we
cannot restrict ourselves by calculation of only two diagrams of Fig.18, but
 we should add also the third diagram.
\section{Lessons for future.}
Inspite of the fact that we failed to construct the effective theory
for high energy interaction starting from the new effective
 degree of freedom -Pomeron we certainly learned a lot about properties of
high energy interaction. In this section I want to summarize those lessons
that we have learned and which could be useful for future.

1. The longitudinal and transverse degrees of freedoms look differently at
high energy and can be treated separately and in different ways.

2.The effective theory at high energy can be reduced to two dimensional theory
and we need to find the energy spectrum of this two dimensional theory to
specify the high energy asymptotic.

3.The major problem of any affective theory at high energy is to find correct
 degree of freedom or in other words the correct effective particle which
is responsible for high energy asymptotic. The concrete realization, namely
Reggeon Calculus failed, because the Pomeron, effective particle of that time,
turns out to be not the correct one to construct theory at high energy
( see  the previous section).

4. The only way to invent correct effective particle is to develop the
consistent approach starting from microscopic theory but not from
phenomenology.

At that time we had no microscopic theory and in the best tradition of high
energy physics the experts  left the field.
Now situation is quite different, we have good microscopic theory ( QCD) and
certainly we have a lot of problems in QCD which  have to be solved. High
 energy asymptotic is only one of many. I think it is time to ask yourselves
why we spend our time and brain trying nevertheless to find the high energy
asymptotic in DIS. Let me give the first answer to this question, namely
 it seems extremely
interesting to go back to old problems to see how far away was our guess from
the systematic approach. It is rather private but strong motivation for many
experts including me. Later we will discuss this problem in more details.

We want to finish our review of the past in high energy physics with a
statement, that inspite of all theoretical inconsistency, all  intristic
contradictions of the Reggeon Calculus, this old fashioned approach is still
the main source of our terminology in high energy theory and still the only
phenomenological approach that we have in hand even now when we discuss so
 called ``soft" processes. It is why the knowledge of this approach is still
 the alpha and omega of high energy strong interaction theory. We hope that we
gave you just right amount of information to move to more theoretical
approach based on QCD.
\newpage
\begin{Huge}
 {\bf TODAY:}
\begin{center}
{\bf  POMERON WITHOUT MAGIC}\\
{\bf IN\,\,\, PERTURBATIVE\,\,\,\,QCD}\\
\end{center}
\end{Huge}
\section{Basics of perturbative QCD (pQCD)}
\subsection{The basics of pQCD for high energy scattering.}
In this subsection we are going to discuss the main ideas of perturbative QCD,
which is our microscopic theory of the strong interaction. The goal of our
 approach is to understand what is the Pomeron in QCD, but we start from the
several things that everybody knows about QCD.

1. If the typical distances ($r$) in our process are small we have
 a natural small
 parameter in QCD, namely the strong coupling constant
$$\as( r) \,\,=\,\,\frac{4 \pi}{b \ln \frac{1}{r^2\Lambda^2}}\,\,=\,\,
\frac{\as(\mu^2)}{1\,+\,\frac{\as(\mu^2) b}{4 \pi} \ln\frac{1}{r^2\mu^2}}
\,\ll\,1
$$
where $ b = 11 - \frac{2}{3} n_f$, $n_f$ is the number of quarks, $\Lambda$
is the QCD confinement scale and $\mu$ is the renormalisation scale.

It means that we can start to calculate the amplitude of our process using the
expansion with respect to small $\as$.

2. The Born approximation of pQCD gives the cross section which is constant
at high energy
$$\sigma_t \,\,\rightarrow |_{s \rightarrow \infty} \,\,Const
$$
We will discuss this property in details in the next section.

3. The more complicated diagrams lead to increase of the total cross
section with energy as we will discuss later
$$\sigma_t \,\,\rightarrow |_{s \rightarrow \infty} \,\, s^{\Delta}
$$
where $\Delta = C \as$.

All the above properties show us that we have a good  chance to get in QCD the
high energy asymptotic which looks like the one Pomeron exchange ( the word
 Pomeron I use here in the traditional sense as the Reggeon with
 intercept close to 1 ) and we are able to examine how good or bad is our
traditional Reggeon - like approach. However, we are going to start with
the brief review what we have learned about pQCD to give you an impression
what kind of theory we have to attack the Pomeron structure.

\subsection{The map of QCD.}

Describing the basic features of QCD,
we will start to answer in this subsection the following questions:
(i) What we have learned about QCD; (ii) What problems
are still unsolved in QCD?; (iii) Can we relate
the kinematical region in which
we have to face these unsolved problems to  collision processes?

Fig.19 displays the `map of QCD'.
Shown are three separate regions, distinguished
by the size of the variables $Q^2$ and $x$,
which will be defined momentarily.
Each region corresponds to quite different
physics and enjoys a different level of understanding.
Before we
discuss these three regions in more detail let us introduce the
necessary notational conventions and definitions.

{\bf Deeply Inelastic  Scattering ( DIS ).}

In Fig.19  $r$ is the distance that can be resolved
by the scattering process under consideration.
{}From the uncertainty principle this distance is
of the order of $\frac{1}{Q}$  where $Q$ is the typical large transverse
momentum in our experiment. E.g.  in deeply inelastic electron
scattering  $Q$ is the transverse momentum of the recoiled electron.
We  recall that deeply inelastic scattering is the
reaction:
$$ e\,\,+\,\,p\,\,\ra \,\,e' \,\,+\,\,anything.$$
Thus, this reaction  acts as a powerful
microscope which
is able to resolve the constituents of a hadron (quarks,
antiquarks and gluons, collectively called partons)
with a transverse size of the order of $\frac{1}{Q}$. The second
kinematical variable that we can introduce for such constituents is the
fraction of energy ($x$)
that a parton carries with respect to the parent hadron.

It should be stressed that the energy in DIS is equal to $s = \frac{Q^2}{x}$.
It means that we are interested in low $x$ DIS if we are going to
 discuss the high energy asymptotic.
If $N(x,Q^2)dx$ is the number of gluons in a small $x$-interval
centered around the value $x$ at scale $Q^2$.
We will show later that the number of
gluons increases in the region of small $x$. All physics in this
kinematical region is strongly connected
to this fact. It is the reason we
concentrate on the discussion of the gluon density here.

The gluon structure function
$x G(x,Q^2)$ tells us what the gluon density at a definite value of
$\ln \frac{1}{x}$ is,
$$ x G (x,Q^2 ) \,\,=\,\,|\frac{d N(x,Q^2)}{d \ln\frac{1}{x}}| \,\,.$$
Let us for convenience introduce the transverse density of gluons
$\rh$,
\beq \label{DEFINITION1}
  \rh(x,Q^2) \,\,=\,\, \frac{x G(x,Q^2)}{\pi \,R^2}\,\,
\eeq
where $R$ is the radius of the hadron.

We plot this on the vertical axis in Fig.19.
In this figure we can see three different regions:

{\bf 1.} {\em The region of small parton density at small distances (low
density (pQCD) region).}

This is the region where we can apply the powerful methods of perturbative
QCD  since the value of running coupling constant $ \alpha_s (1/r^2)$ is
small ($\alpha_s (1/r^2)\,\, \ll \,\,1$). During two
 decades remarkable
 theoretical progress has been achieved here (GLAP evolution equation
\cite{GLAP},
 gluon bremshtrahlung for jet decay \cite{DOMUBOOK} , factorization theorem
 (J.Collins,D.Soper and G.Sterman (1983) \cite{FACTEO})
and the main property of ``hard" processes has been experimentally
 confirmed  at LEP and at the Tevatron.

{\bf 2.} {\em The region of large distances (npQCD region).}

Here we have to deal with the confinement problems of QCD, since
$\alpha_s (1/r^2) \,\gg \,1$. In this kinematical
region we need to use nonperturbative methods such as lattice calculation
\cite{KROMA} or QCG Sum Rules \cite{SHI}. The progress here is remarkable but
all developed methods cannot yet be applied to scattering processes.

{\bf 3} {\em The region of small distances but high parton density
(hdQCD region).}

 Here we have a unique situation in which
the coupling constant $\alpha_s $ is still small but the density is so
large that we cannot use the usual methods of
perturbation theory.

 In essence the
theoretical problem here is also a nonperturbative one but the origin of
the nonperturbative effects here is quite different from that in the
previous region.
Here we face the situation where we have to
develop new methods that let
deal with a dense relativistic system of
gluons in a nonequilibrium state.
Unfortunately we are only at the beginning of this road.

 Fortunately, we can control theoretically this dense system of partons
in some transition region on the border of the pQCD and
hdQCD regions and here we can study this remarkable system of partons in great
detail. Thus the right strategy is to approach this interesting kinematic
 region from the low density pQCD region.

In my opinion the  interest to high energy scattering was revived only because
 moving from pQCD region to hdQCD region we can open new window to
 study  a nonperturbative nature of QCD.
\subsection{What fundamental problems could be solved?}
This is why we want to list here the fundamental problems that we hope
 to solve penetrating high density QCD region.
We hope:

1. to specify the kinematical region in which we can trust pQCD
 (GLAP evolution equation, gluon bremstrahlung, factorization theorem ...);

2. to find new collective phenomena for nonabelian
 theories such as QCD ;

3.  to find the analytic solution of hd QCD which is nonperturbative but
 looks simpler than np QCD since $\alpha_s \ll 1$ here;

4. to develop methods with which  build an effective theory for hd QCD.

We need the effective theory because  we can use the Lagrangian of such a
 theory for exact calculation on lattice, for example, or even we can try to
 solve problem analytically.
\subsection{How to penetrate the high density QCD region.}
Access to this interesting kinematical
region is actually easily achieved in our scattering processes.
We know at least three ways
to prepare a large density system of partons.

  {\bf 1.}
The first is given by nature,
which supplies us with large and heavy nuclei.
In ion-ion collisions we can already reach a very
high density of partons
at not so high energies,  because the partons from different
nucleons in a nucleus are freed.

{\bf 2.}
The second relates to
hard processes in hadron-hadron collisions or in deep inelastic
scattering. These also give us access to a high density of partons because we
expect a substantial increase in the
parton density in the region of small Bjorken $x$. The experimental
 data from HERA show the significant increase of the deep inelastic
structure function:
$$ F_2(Q^2,x_B) \,\propto\, ( \frac{1}{x_B}
)^{0.33}  \,\,at \,\,Q^2\,\sim\,10 \,GeV^2.$$

{\bf 3.} The third is to measure the event with sufficiently large multiplicity
of produced particles, larger than the multiplicity in the typical inelastic
(bias) event.

Thus we can formulate the ideal experiment for search of high density partonic
system:

{\it`` The deeply inelastic scattering with nucleus at low $x$
with special selection of events with large multiplicity of produced particles.
" }

\subsection{Method  of pQCD( two diabols that we are fighting with).}
 Accordingly to our strategy we start with perturbative QCD in that region of
Fig.19  where the density and coupling QCD constant ($\alpha_s$) are small.
Each observable (i.e gluon structure function) could be written in pQCD as
following series:
\beq
xG(x,Q^2)\,\,=\,\,\Sigma_{n = 0} \,C_n ( \alpha_s)^n\cdot(L^n + a_{n - 1}
L^{n - 1} ... a_0)\
,\,.
\eeq
We have two big problems with this perturbative series which are two our
biggest
 enemies:

{\bf 1.} The natural small parameter $\alpha_s$ is compensated by large log
(L).
The value of L depends on the process and kinematic region. For example in
deeply inelastic scattering (DIS):
$$ \L \,\,=\,\,\log Q^2 \,\,\,\,\,\,at\,\,\,\, Q^2 \gg Q^2_0\,\,but
\,\,x\,\sim\,1
$$
$$
\L\,\,=\,\,\log(1/x)\,\,\,\,\,\,\,at\,\,\,\, Q^2\,\sim\,Q^2_0
\,\,and\,x\,\rightarrow\,0
$$
$$
\L\,\,=\,\log Q^2\,\cdot\,\log(1/x)\,\,at\,\,\,\, Q^2\,\gg Q^2_0 \,\,and
\,x\,\rightarrow
\,0
$$
$$
\L\,\,=\,\,\log(1 - x)\,\, \,\, at\,\,\,\, Q^2\,\sim\,\,Q^2_0 \,\,and\,\, x\,
\rightarrow\,1
$$
Of course it is not the full list of scales. The only that I would like to
 emphasize that $L$ depends on the kinematic region. Thus to calculate
 $xG(x,Q^2)$ one cannot calculate only the Born Approximation but has to
calculate the huge number of Feynman diagrams.

{\bf 2.}
$$
\,\,\,\,\,C_N\,\,\rightarrow\,\,n! \,\,\,\,\,at\,\,\,\, n\,\,\gg\,\,1
$$
It means that we are dealing with asymptotic series and we do not know the
 general rules what to do with such series. There is only one rule, namely to
 find the analytic function which has the same perturbative series. Sometimes
but very rarely we can find such analytic function. In this case this is the
 exact solution of our  problem. Mainly we develop some general approach
based on Leading Log Approximation (LLA). The idea is simple. Let us find
the analytic function that sums the series:
\beq
xG(x,Q^2)_{LLA}\,\,=\,\,\Sigma_{n=0} C_n\, (\alpha_s \cdot L )^n\,\,.
\eeq
Ussualy we can write the equation for function $xG(x,Q^2)_{LLA}$. The most
famous one, the GLAP evolution equation \cite{GLAP}, sums eq.(2) if $L$=
$\log Q^2$. The BFKL \cite{BFKL} equation gives the answer for eq.(2) in the
case when $\L$=$\log(1/x)$. Using the solution of the LLA equation we built
the ratio:
\beq
R(x,Q^2)\,\,\frac{xG(x,Q^2)}{xG(x,Q^2)_{LLA}}\,\,=\,\,\Sigma_{n=0} r^n\,=
\Sigma_{n=0} c_n\cdot(L^{n-1} + a_{n-1}L^{n - 2} + ... a_0)\,\,.
\eeq
This ratio is also asymptotic series but what we are doing we calculate
this series term by term. Our hope is that the value of the next term will be
 smaller then the previous one $( \frac{r_{n} }{r_{n - 1}} \ll 1 )$ for
 sufficiently large $n$. However we know that at some value of $n=N$
$\frac{r_{N}}{r_{N-1}}\,\sim \,1$. The only that we can say about such
situation
 in  general that our calculation has intristic theoretical accuracy and the
 result of calculation should be presented in the form
\beq
R(x,Q^2)\,\,=\,\,\Sigma^{n = N-1}_{n=0} r_{n} \,\pm r_{N}\,\,.
\eeq
How big is the value of $N$ depends mostly on how well we chose the LLA and how
well we established the value of scale $L$ in the process of interest.
\subsection{ Present Theoretical Status = Regeneration of Reggeon Calculus.}
What we are doing now approaching the high density QCD domain is really the
regeneration of the old idea of Reggeon Calculus, namely we
reduce the complicated problem of quark and gluon interaction in the dense
parton system to the interaction of our ``building bricks", so called Pomerons
( see Fig. 20, where the structure of our approach is shown). It looks like in
old, good time of Reggeon Dominance but we have two new and very important
ingredients:

{\bf 1.}
  The QCD Pomeron is not an  invention but naturally appears in perturbative
QCD
 in the   leading $log (1/x)$
 approximation
(LL (x)A) to the scattering amplitude at high energy
( the Balitski - Fadin -Kuraev - Lipatov (BFKL ) equation
\cite{BFKL} ). It means that in a restricted kinematical region
the BFKL Pomeron describes the high energy interaction within a certain
  guaranteed theoretical accuracy. This fact makes it unavoidable that
one should build an effective theory starting with the BFKL Pomeron.

{\bf 2.}
The vertices of the Pomeron interactions are not phenomenological parameters
but can be calculated in perturbative QCD. It turns out that only two of them
are essential in the vicinity of the border between pQCD and hdQCD regions in
 Fig.19, namely triple Pomeron vertex ($\gamma$) and rescattering of two
 Pomerons ($\lambda$) (see Fig.20).

We are going to discuss in detail the structure of the BFKL Pomeron as well as
the theory of the shadowing corrections. The process which we consider is the
 deeply inelastic scattering in  which all high energy phenomena occur at
small distances where we can trust the pQCD calculation.

\section{ Pomeron in pQCD (the BFKL Pomeron)}

The BFKL equation  was derived in so called Leading Log (1/x)
Approximation, in
which we would like to keep the contribution of the order of $( \as log
(1/x))^n$ and neglect all other contributions, even of the order of $\as\,\,
log (Q^2/Q^2_0)$. So the set of parameters in  LL(1/x)A is obvious:
$$
\as\,\, log \frac{1}{x} \,\,\sim\,\,1\,\,;
$$
\beq
\as\,\, log \frac{Q^2}{Q^2_0} \,\,< \,\,1\,\,;
\eeq
$$
\as\,\,\,\ll\,\,1\,\,;
$$
Let us consider the simplest process: the quark - quark scattering at high
energy at zero  momentum transfer . All problems of infrared divergency in
such a process is irrelevant since they are canceled in the scattering of two
colorless hadrons . We will show this fact using the example of
 the first diagram (the Born Approximation (BA)).
\subsection{The Born Approximation}
In the Born Approximation the only diagram of Fig. 21(a)  contributes to the
imaginary part of the scattering amplitude (A). It is easy to understand
that the result of calculation of this diagram gives \footnote{Let me recall
you that we are keeping the coupling constant fixed.}:
\beq \label{BA}
 2 \Im{A^{BA}(s,t = 0 )}\,\,=\,\,\int \frac{ d^2 k_t}{ ( 2 \pi)^2} |\,\, M ( 2
\ra 2;\, \as\,\, |\,\, s, t = - k^2_t)\,\, |^2
\,\,=\,\,s \frac{C^2_2 \as^2}{ N^2 - 1} \,\int\,\frac{ d^2 k_t}{ k^4_t}\,\,,
\eeq
where $ M (2 \ra 2;\, \as\, |\, s, t = - k^2_t ) $ denotes the amplitude in
the lowest order of $\as$ for
quark - quark scattering at transfer momentum $ t= - k^2_t $ through one gluon
exchange ( see Fig.21(a) ),
N is the number of colours and $C_2 \,=\,\frac{ N^2 - 1}{ 2 N} $. One can
recognize the Low - Nussinov mechanism \cite{LONU} of high energy
interaction in this simple example.

Let us consider the BA in more detail for the scattering of two mesons
( $a$ and $b$ ), assuming that each of them  consists of two heavy quarks.
Such a system has sufficiently small size, namely
$$r\,\,\propto\,\,\frac{1}{M_Q \as}$$
and can be treated in pQCD.

Let me remind you why the size is small. The total energy of two quarks in
 Coulomb potential is equal:
$$
E\,\,=\,\,\frac{p^2}{M_Q}\,\,-\,\,\frac{\as}{r}
$$
If we remember the uncertainty principle
$$
\Delta p \Delta r \,\simeq 1\,\,\,\,or\,\,\,\,p\,\,\simeq\,\,\frac{1}{r}
$$
we get for the system with $E \rightarrow 0 $ that size which we have
 mentioned before.

 Repeating the calculations we get the
same answer as for quark - quark scattering, namely
\beq
\sigma^{BA}\,\,=\,\,\as^2\,\frac{2}{9}\cdot 4\cdot n_a n_b \int \frac{d^2 k_t}{
( k - \frac{q}{2} )^2_t \,( k + \frac{q}{2} )^2_t}\,\cdot \,\Phi_a(k,q)
\,\Phi_b (k,q)
\eeq
where $n_a (n_b)$ is the number of quarks in the meson $a$($b$), the factor
$\frac{2}{9}$ is the result of averaging over colours  of the initial quarks
and $\Phi (k,q)$ is the amplitude of the emission of two gluons by a meson.
$$\bullet$$
{\bf Problem 15:} Show that the emission amplitude $\Phi$ can be expressed
in terms of meson formfactors if we can treat the heavy quarks as
nonrelativistic particles (see Fig.22(c)) :
$$
G(q^2)_a\,\,=\,\,\int d p \Psi_a (p) \,\Psi^{*}_a (p - q ) ;
$$
$$
\Phi_a (k,q )\,\,=\,\,\int  d p \Psi_a (p) \,\Psi^{*}_a (p - q )\,\,-\,\,
\int d p \Psi_a (p + \frac{q}{2}  + k ) \,\Psi^{*}_a (p  + \frac{q}{2} - k )
\,\,=\,\,G_a (q^2) - G_a ( 4 k^2 ).
$$
$$\bullet$$
Finally, we have
\beq
\sigma^{BA}\,\,=\,\,\as^2\,\frac{2}{9}\cdot 4\cdot n_a n_b \int \frac{d^2 k_t}{
k^4_t}\,\cdot \,
[\,G_a (0) - G_a ( 4 k^2 )\,]
\,[\,G_b (0) - G_b ( 4 k^2 )\,]
\eeq
Now let us ask the question what kinematic region of integration over $k_t$
gives the biggest contribution to the total cross section
if two mesons have
different sizes, let say $r_a \,\gg\,r_b$. We can study three regions
 of integration:

1. $\lambda\, >\, r_a$, where $\lambda$ is the wave length of the gluon
( $\lambda\,=\,\frac{1}{k_t}$ ).

In this case $k_t r_a \,<\,1 $ as well as $k_t \,r_b \,<\,1$ and
$$G_a (0) - G_a ( 4 k^2 )\,\rightarrow \,\frac{4}{6} r^2_a k^2_t$$
$$G_b (0) - G_b ( 4 k^2 )\, \rightarrow \,\frac{4}{6} r^2_b k^2_t$$
The integral over $k_t$ gives you the answer
$$
\sigma^{BA} \,\simeq \,r^2_b
$$

2.$  r_a\,>\,\lambda\,>r_b$

In this case $k_t r_a \,>\,1 $ as well as $k_t \,r_b \,<\,1$ and
$$G_a (0) - G_a ( 4 k^2 )\,\rightarrow \, 1$$
$$G_b (0) - G_b ( 4 k^2 )\, \rightarrow \,\frac{4}{6} r^2_b k^2_t$$
The integral over $k_t$ gives you the answer
\beq
\sigma^{BA}\,\,=\,\,\as^2\,\frac{2\pi}{9}\cdot 4\cdot n_a n_b
\,\cdot\,\frac{2}{3}\,
\cdot\,r^2_b \,\ln \frac{r^2_a}{r^2_b}
\eeq

3. $r_b\,>\,\lambda$

It means that
 $k_t r_a \,>\,1 $ as well as $k_t \,r_b \,>\,1$ and
$$G_a (0) - G_a ( 4 k^2 )\,\rightarrow \, 1$$
$$G_b (0) - G_b ( 4 k^2 )\, \rightarrow \,1$$
The integral over $k_t$ gives you  the same answer as in the first kinematic
 region, namely:
$$
\sigma^{BA} \,\simeq \,r^2_b
$$
This example is very instructive, because our standard process - DIS which
we can treat using pQCD and in DIS the typical size of our probe - virtual
 photon is the smallest one in the process and is of the order of
$\frac{1}{Q}$.
It means that the BA gives for the DIS the total cross section which
is proportional $\frac{1}{Q^2} \ln (Q^2R^2)$ where $R$ is the
 size of the target. It should be stressed as has been mentioned that the
total cross section does not depend on energy in BA.
\subsection{The next  order of $\as$  approximation.}
In the next order we have to consider a larger  number of the diagrams, but
 we can write down the answer in the following general form:
\beq \label{NEBA}
2 \Im{A^{NBA} ( s,t=0)} \,\,=\,\,
\eeq
$$
\int ( 2 \pi )^4 \delta^{(4)}( {\bf p_1 + p_2 -
p'_1 - p'_2 - q } ) |\,\, M ( 2 \ra 3; g^3)\,\,|^2\,\, \Pi^{i = 3}_{i = 1}
\frac{d^3 p'_i}{ ( 2 \pi )^3  2 E'_i}  \,\,+$$
$$
 \int ( 2 \pi )^4 \delta^{(4)} ( {\bf p_1 +
p_1 - p'_1 - p'_2})\cdot 2\, \Re{
 M ( 2 \ra 2 ; \as^2 ) \cdot M^{*} (2 \ra 2 ; \as)}
 \Pi^{ i = 2}_{i = 1}\,\, \frac{ d^3 p'_i}{ ( 2 \pi)^3 2 E'_i}\,\,,
$$
where $g$ is coupling constant of QCD ( $\as = \frac{g^2}{4 \pi}$);
$ M( 2 \ra 3; g^3)$ is the amplitude for the production of the extra gluon
in the Born Approximation, and  is given by the set of Feynman diagram in
Fig.22  while $ M ( 2 \ra 2, \as^2 ) $ is the amplitude of the elastic
scattering in the next to leading Born Approximation ( see Fig. 23) at the
momentum transfer $k_t$.

Two terms in \eq{NEBA} have different physical meaning: the first one
describes the emission of the additional gluon in the final state of our
reaction,  while the second term is the virtual correction to the Born
Approximation due to the emission of the additional gluon. It corresponds to
the same two particle final state, and describes the fact that due to
emission the probability to detect this final state becomes smaller ( we
will see later that the sign of the second term is negative).

In the both terms of \eq{NEBA} we can integrate over ${ \bf \vec{p_3}}'$ as
well as
over the longitudinal component of ${\bf {\vec  p'_1}}$ ($p'_{1L}$).
Finally, we  rewrite the phase space in the following way:
$$
\int ( 2 \pi )^4 \delta^{(4)}( {\bf p_1 + p_2 -
p'_1 - p'_2 - q }\,\, \Pi^{i = 3}_{i = 1}\frac{d^3 p'_i}{ ( 2 \pi )^3  2
E'_i}\,\,=\,\,\frac{1}{ 4 \pi}\cdot\frac{1}{s} \cdot \int^{1}_{
x_{min} = \frac{m^2}{s}}
\frac{d x'_3}{x'_3} \int \frac{ d^2 p'_{1t} d^2 p'_{3t}}{ ( 2\pi )^4}\,\,;
$$
\beq \label{PS}
\int ( 2 \pi )^4 \delta^{(4)} ( {\bf p_1 +
p_1 - p'_1 - p'_2})\cdot\Pi^{ i = 2}_{i = 1}\,\,
 \frac{ d^3 p'_i}{ ( 2 \pi)^3 2 E'_i}\,\,=\,\, \frac{1}{ s} \cdot \int
\frac{d^2 p'_{1t}}{ ( 2 \pi)^2}\,\,;
\eeq
where $x$ is the fraction of the longitudinal momentum carried by particle.
the value of $x_{min}$ depends on the  reaction. For example in the deeply
inelastic scattering $x_{min}\,=\,x_{B}\,=\,\frac{|Q^2|}{s}$. In the case of
the quark scattering $x_{min} \,=\,\frac{m^2_t}{s}$ where $m_t$ is the
transverse mass of produced quark.

 From \eq{PS} one can see the origin of the $log (1/x_{min})$ contribution:
it stems from the phase space integration, if $ M ( 2\,\ra\,3 )$ does not
go
to zero at $ x_3 \,\ra \, 0 $. To sum the diagrams of Fig. 22  in this
limit we can use two tricks. The first one is  for each $t$-
channel gluon one can rewrite the numerator of the gluon propagator at high
energy in the following way:
\beq \label{GLUON}
g_{\m \n}\,\,=\,\,\frac{ p_{1 \m} p_{2 \n} \,\,+\,\,p_{2 \m} p_{1 \n}}{
p_1\cdot p_2}\,\,+\,\, O
(\,\, \frac{m^2}{ s}\,\,)
\eeq
The second trick is based on the gauge invariance of the QCD. We look on
the subset of the diagrams of Fig. 22  pictured in Fig. 24  as the
amplitude of the interaction of gluon $ k$ with the quark $p_2$ ( see
Fig.24 ).  Since all particles in amplitude $ M_{\n}$ except gluon $k$
are on the mass shell, the gauge invariance leads to the relationship:
\beq \label{GAUGETRICK}
k_{\n} M_{\n} \,\,=\,\,0\,\,.
\eeq
Using Sudakov variables \cite{SUDAKOV} we can expand  vector $k$ as
$$
k_{\n}\,\,=\,\,\a_{k} \,p_{1\n} \,\,+\,\,\b_{k}\,p_{2\m} \,\,+\,\,k_{t \m}
$$
and  rewrite \eq{GAUGETRICK} in the form:
\beq \label{GATR}
\(\,\,\a_{k} \,p_{1\n} \,\,+\,\,\b_{k}\,p_{2\m} \,\,+\,\,k_{t
\m}\,\,\)\,\,M_{\n}\,\,=\,\,0\,\,.
\eeq
 We  note that all particle inside $ M_{\n}$ have a large component of
their momentum on $p_2$. It means that we can neglect the projection of
vector $M_{\n}$ on $p_1$,  or in other words
$$
M_{\n}\,\,=\,\, M^{(1)} p_{1 \n} \,\,+\,\,M^{(2)} p_{2 \n}
\,\,+\,\,M^{(t)}_{\n}
$$
 and $ M^{(1)}\,\ll\,M^{(2)} $.
 Thus we can conclude from \eq{GATR} that
\beq \label{TRPO}
 p_{1 \n} M_{\n}\,\,=\,\,-\,\frac{k_{t \n} M_{\n}}{\a_k}\,\,.
\eeq
 Using both tricks of \eq{GLUON} and \eq{TRPO} one can easily see that only
diagram of Fig. 24(1) contributes in LL(log(1/x)A. Indeed, let us
consider for example the diagram of Fig. 24(2), the dominator of the quark
propagator $( p_2 + k )^2 $ is equal to
$$
( p_2\,+\,k )^2\,\,=\,\,\a_k s \,\,+\,\,k^2_t\,\,=\,\,-\,\frac{p'^2_{3t}}{x_3}
\,+\,k^2_t\,\,\sim\,\,-\,\frac{p'^2_{3t}}{x_3}\,\,.
$$
Since due to \eq{TRPO} the polarization of gluon $k$ is transverse we cannot
compensate the smallness this diagram at $x_3 \,\ra\,0$. Using the same
tricks with the upper parts of the diagrams of Fig. 22 we arrive at the
conclusion that that the set of the diagrams of Fig. 22 degenerates into
one diagram of Fig. 25 with specific vertex for gluon emission:
\beq \label{VERTEX}
\Gamma_{\sigma}\,\,=\,\, i\,g\,f_{a b c}\cdot \frac{2 \, k_{t \m}\,k'_{t \n}}
{ \a_k \b_{k'} s}\, \gamma_{\m \sigma \n}\,\,,
\eeq
 where $\gamma_{\m \sigma \n}$ is  given by the usual Feynman rules for QCD.

Substituting  \eq{VERTEX} into the first term of \eq{NEBA} we  get the
contribution of the emission of one additional gluon to the next to Born
Approximation in the form:
\beq \label{EMINBA}
\Im{A^{NBA}_{emission}}\,\,\,\,=\,\,s\,\frac{\as^2\,C^2_2}{ N^2\,-\,1}
\int^y_0 d y' \,\int\,\frac{d^2 k_t}{\pi\, k^4}\,\cdot \frac{ N \as}{\pi}
\cdot  K_{emission} (
k_t, k'_t )\cdot \,\frac{d^2 k'_t}{\pi\, k'^4} \,\,,
\eeq
where $ y'\,=\,log (1/x_3)$, $y \,=\,log (1 / x_{min}$  and
 the kernel $K (k_t, k'_t )$ is equal to
\beq \label{KERNEM}
K_{emission} (k_t, k'_t)\,\,=\,\,\frac{ k^2_t\, k'^2_t}{
(\,k_t\,\,-\,\,k'_t\,)^2} \eeq.

To calculate the virtual correction in the next to Born Approximation
 ($ M (\,2\,\ra\,2\,;\as^2\,)$ )  we have
to estimate the contribution of  the set of diagrams of Fig. 23.
The log (1/x) contribution is hidden in the real part of the amplitude
$ M (\,2\,\ra\,2\,;\as^2\,)$, and easiest way to extract this log is to
use the dispersion relation:
\beq \label{DISRE}
M(\,2\,\ra\,2\,;\,\as^2\,)\,\,=\,\,\frac{1}{\pi} \cdot \{\,\,\int
\frac{\Im{M(\,2\,\ra\,2\,;\,\as^2\,)}_s}{ s'\,\,-\,\,s} \,d\,s' \,\,+\,\,
\frac{\Im{M(\,2\,\ra\,2\,;\,\as^2\,)}_u}{ u'\,\,-\,\,u} \,d\,u'\,\,\}\,\,.
\eeq
We calculate  $\Im{M(\,2\,\ra\,2\,;\,\as^2\,)}_s$ and
$ \Im{M(\,2\,\ra\,2\,;\,\as^2\,)}_u$ using unitarity ( see \eq{BA}):
\beq \label{UNITARITY}
 \Im{M(\,2\,\ra\,2\,;\,\as^2\,\,| \,t\,=\, - k^2_t\,)}_s\,\, =
\eeq
$$
\,\,\int \frac{ d^2 k'_t}{ ( 2 \pi)^2} |\,\,
\Re{M ( 2 \ra 2;\, \as\,\, |\,\, s, t = - k'^2_t)\,\,
M ( 2 \ra 2;\, \as\,\, |\,\, s, t = - ( k_t - k'_t )^2\,)}\,\,.
$$
The difference between \eq{UNITARITY} and \eq{BA} is that the Born amplitude
for one gluon exchange enters these two equations at different values of
momentum  transferred $t$. The explicit calculations give:
\beq \label{RENBA}
 \Im{M(\,2\,\ra\,2\,;\,\as^2\,\,| \,t\,=\, - k^2_t\,)}_s \,\,=
\,\,s\,\cdot\,C_s\,\pi\,\Sigma ( k^2_t );
\eeq
$$
\Im{M(\,2\,\ra\,2\,;\,\as^2\,\,| \,t\,=\, - k^2_t\,)}_u \,\,=
\,\,s\,\cdot\,C_u\,\pi \,\Sigma ( k^2_t );
$$
$$
\Sigma ( k^2_t ) \,\,=\,\, \frac{4 \as }{\pi^2}\,\int \frac{d^2 k'_t}{ (
\,k_t - k'_t \,)^2\,k'^2_t}\,\,;
$$
Using the dispersion relation of \eq{DISRE} one can reconstruct the real part
of the amplitude and the answer is
\beq \label{RE}
\Re{M(\,2\,\ra\,2\,;\,\as^2\,\,| \,t\,=\, -
k^2_t\,)}\,\,=\,\, s\, (\, C_u \,-\, C_s\,)\cdot \Sigma ( k^2_t ) \cdot log s
\eeq
The colour coefficients have a very famous relation between them ( see Fig.
26 ) which gives, for the difference of the colour coefficients in \eq{RE}
the same colour structure as for the diagram of Fig. 23 ( 3 ). Thus
$\Re{M(\,2\,\ra\,2\,;\,\as^2\,\,| \,t\,=\, - k^2_t\,)}$
 has the same colour structure as one gluon exchange in the Born Approximation.
This fact makes it  possible to rewrite the second term in \eq{NEBA} as the
correction to the gluon trajectory, e.g. instead of gluon with propagator
$\frac{1}{k^2 }\cdot s $ we can introduce the new propagator
\beq \label{REGGE}
\frac{1}{k^2}\cdot s^{\a^{G}( k^2)}
\eeq
$$
\a^{G}(k^2)\,\,=\,\, 1\,\,-\,\,\frac{\as N}{ \pi^2}\cdot \int \frac{ k^2_t
d^2 k'_t}{ (\,k_t\,-\,k'_t\,)^2 \, k'^2_t}\,\,=\,\,1\,\,-\,\,\frac{\as N}{
 2 \pi^2}\cdot \int \frac{ k^2_t
d^2 k'_t}{[\, (\,k_t\,-\,k'_t\,)^2 \,+\, k'^2_t\,] k'^2_t}
$$
The answer for the second term in \eq{NEBA} can be written in the form
\beq \label{VIRTUAL}
\Im{ A^{NBA}_{virtual}}\,\,=\,\,s\,\frac{\as^2 C^2_2}{ N^2 \,-\,1}\,
\int  \(\, \a^{G} ( k^2 )\,\,-\,\,1\,\)\cdot 2 \cdot \frac{ d^2 k_t}{
k^4_t}\,\,.
\eeq
We can get the full answer for the amplitude in the next to the Born
Approximation ($\as^3$ by summing \eq{EMINBA} and \eq{VIRTUAL} and it can be
written in the form:
\beq \label{FINNBA}
\Im{A^{NBA} ( s, t = 0 )} \,\,=\,\,s\,\frac{\as^2\,C^2_2}{
N^2\,-\,1}\,\,\int^y_0 d y' \int \,\frac{d^2 k_t}{ k^2_t}\cdot\frac{ \as
N}{\pi^2} \cdot K ( k_t, k'_t ) \cdot \frac{d^2 k'_t}{ k'^2_t}
\eeq
where $ y \,=\,log (1/x_{min}$ and
\beq \label{LIKER}
K ( k_t, k'_t ) \cdot \frac{1}{ k'^2_t} \,\,=\,\,
\frac{1}{ (\,k_t\,-\,k'_t\,)^2}\cdot\frac{1}{k'^2_t}\,\,-
\,\,\frac{k^2_t}{(\,k_t\,-\,k'_t\,)^2\,[\, (\, k_t\,-\,k'_t\,)^2\,+\,k'^2_t\,]}
\cdot\frac{1}{k^2_t}\,\,.
\eeq
Using \eqs{FINNBA}{LIKER} we can introduce function $\p ( k^2_t)$ and rewrite
the
total cross section for quark - quark scattering in the form:
\beq \label{QUARKX}
\s_{qq}\,\,=\,\,\frac{\as C_2}{ N^2\,-\,1}\cdot \int \p ( y, k^2_t) \cdot
\frac{ d k^2_t}{ k^2_t}\,\,.
\eeq
For $\p$ \eq{FINNBA} gives the equation:
\beq \label{LIEQNBA}
\p^{(2)} ( y, k^2_t )\,\,=\,\,\frac{ \as N}{ \pi^2} \cdot \int^y_0 d y' \int
d^2 k'_t
K (k_t, k'_t) \,\,\p^{(1)} ( y', k'^2_t )\,\,.
\eeq
where
\beq \label{LIKERNEL}
K ( k_t, k'_t ) \p ( k^2_t ) \,\,=\,\,
\frac{1}{ (\,k_t\,-\,k'_t\,)^2}\cdot \p ( k'^2_t ) \,\,-
\,\,\frac{k^2_t}{(\,k_t\,-\,k'_t\,)^2\,[\, (\, k_t\,-\,k'_t\,)^2\,+\,k'^2_t\,]}
\cdot \p ( k^2_t )\,\,.
\eeq
and
\beq \label{INCONG}
\p^{(1)}\,\,=\,\, \frac{\as C_2}{ k^2_t}\,\,.
\eeq
\subsection{The main property of the BFKL equation.}
 From the simplest calculation in $\as^3$ order one can guess the BFKL equation
\cite{BFKL} which looks as follows:
\beq \label{BFKL}
\frac{d \p (\, y\,=\,log (1/x ), k^2_t\, )}{ d y} \,\,=\,\,\frac{\as
N}{\pi^2} \cdot \int \, K (\,k_t, k'_t\,)\,\p (\,y, k'^2_t\,)\,d^2 k'_t\,\,,
\eeq
where kernel $ K (k_t, k'_t) $ is defined by \eq{LIKERNEL}.
This equation sums the $ ( \as log (1/x ) )^n $ contributions and has
"ladder" - like structure   ( see Fig. 27 ). However, such "ladder"
diagrams are only an effective representation of the whole huge set of the
Feynman diagrams, as explained in the simplest example of
the previous subsection. The first part of the kernel $ K (k_t, k'_t )$
describes the emission of new gluon, but with the vertex which differs from
the vertex in the Feynman diagram, while the second one is related to the
reggeization of all t-channel gluons in the "ladder".

 The solution of the BFKL equation has been given in ref. \cite{BFKL} and we
we would like to recall some main properties of this solution.

\subsubsection{Eigenfunctions of the BFKL equation.}
The eigenfunction of the kernel $K(k_t,k'_t)$ is $\p_f\,\,=\,\,(k^2_t)^{f
- 1}$. Indeed after sufficiently long algebra we can see that
\beq \label{EIGENVALUE}
\frac{1}{\pi} \int d^2 k'_t  K (k_t,k'_t)\,\, \,\p_f ( k'^2_t )\,\,=\,\,\c
( f) \,\,\,\p_f ( k^2_t) \eeq
where
\beq \label{CHI}
\c (f)\,\,=\,\,2 \,\Psi (1) \,\,-\,\,\Psi ( f ) \,\,-\,\,\Psi ( 1 - f )\,\,
\eeq
and
$$\Psi ( f )\,\,=\,\,\frac{ d\,\, \ln \Gamma ( f )}{ d\,\, f}\,\,,$$
$\Gamma ( f ) $ is the Euler gamma function.

\subsubsection{ The general solution of the BFKL equation.}
 From \eq{EIGENVALUE} we can easily  find the general solution of the
BFKL equation using double Mellin transform:
\beq \label{MELLIN}
\p (y, k^2_t)\,\,= \,\,\int \,\,\frac{d \omega}{ 2 \pi \,i}
e^{\omega\,y} \p ( \omega, k^2_t)\,\,=\,\,\int \frac{d
\omega d f}{( 2 \pi \,i)^2} e^{\omega\,y} \p_f ( k^2_t )\, C ( \omega, f )
\eeq
where the contours of integration over $\omega$ and $f$ are situated to the
right of all singularities of $\p ( f ) $ and $ C ( \omega, f ) $.
For $C( \omega, f )$ the equation reads
\beq \label{OMEGALI}
\omega\,C ( \omega, f )\,\,=\,\,\frac{\as N }{ \pi}\,\c ( f ) \, C ( \omega,
f )\,\,.
\eeq
Finally, the general solution is
\beq \label{GENSOL}
\p (y, k^2_t ) \,\,=\,\,\frac{1}{  2 \pi \,i }\,\int \,d f\,e^{\,\frac{\as
N}{\pi}\,
\c ( f ) y \,+\, ( f - 1 )\,r}  \tilde \p ( f )
\eeq
where $\tilde \p ( f ) $ should be calculated from the initial condition at
$ y\,=\,y_0 $ and $ r\,=\, ln \frac{k^2_t}{q^2_0}$ ( $q^2_0$ is the value
of virtuality from which we are able to apply perturbative QCD ) .

\subsubsection{Anomalous dimension from the BFKL equation.}
 We can solve \eq{OMEGALI} in a different way and find $ f \,=\,\g (\omega)$.
  $\g (\omega)$ is the anomalous dimension in LL(log (1/x)A
\footnote{From \eq{MELLIN} one can notice that  moment variable N defined
such that N = $\omega$ + 1.}
and for $\g( \omega ) $ we have the following series \cite{LIANDI}
\beq \label{LIANDI}
\g ( \omega ) \,\,=\,\,\frac{\as N }{\pi}\cdot \frac{1}{\omega} \,\,+\,\,
\frac{2  \as^4 N^4 \zeta ( 3 )}{\pi^4} \cdot\frac{1}{\omega^4}\,\,+\,\,O
( \frac{\as^5}{ \omega^5} )
\eeq
$$\bullet$$
{\bf Problem 16:} Get eq.(81) from eq.(79), expanding it with respect to
$f = \gamma (\omega)$.
$$\bullet$$
The first term in \eq{LIANDI} is the anomalous dimension of the GLAP equation
\cite{GLAP} in leading order of $\as$ at $\omega \,\ra\,0$, which gives the
solution for the structure function at $ x\,\ra\,0 $ and corresponds to so
called double log approximation of perturbative QCD ( DLA).The  DLA sums
 the contributions of the order $ ( \as \,log (1/x)\,log (Q^2/q^2_0) )^n$
 in the perturbative series of eq.(45).

However, we would like to stress that \eq{LIANDI} is valid only at fixed
 $\as$ while the anomalous dimension in the GLAP equation can be calculated
for running $\as$. It means that we have to introduce the running $\as$ in
the BFKL equation to achieve a matching with the GLAP equation in the region
where $\omega \,\ll\,1$ and $ \frac{\as}{\omega} \,< \, 1$.

The second remark is the fact that we can trust the series of \eq{LIANDI}
only for the value of $\oa \,\gg \,\ol$ , where
\beq \label{OMEGALIP}
\ol\,\,=\,\,\frac{\as \,N}{\pi}\cdot \c (\,\frac{1}{2}\,)\,\,=\,\,\frac{ 4 N
\,ln 2\,\as}{\pi}\,\,.
\eeq
In vicinity $ \oa\,\ra \ol $ we have the following expression for $\gamma(
\oa ) $:
\beq \label{GAMMAATOL}
\gamma ( \oa )\,\,=\,\,\frac{1}{2}\,\,+\,\,\sqrt{\frac{
\oa\,\,-\,\,\ol}{\frac{\as N}{\pi} \,14 \zeta ( 3 )}}\,\,.
\eeq
$$\bullet$$
{\bf Problem 17:} Get eq.(83) front eq. (79), using the property of $\Psi (f)$.
$$\bullet$$
Substituting \eq{GAMMAATOL} in \eq{MELLIN} we have
\beq \label{SOLATOL}
\p (y, k^2_t)\,\,= \,\,\int \,\,\frac{d \omega}{ 2 \pi \,i}\,\,
e^{\omega\,y} \p ( \omega, k^2_t)\,\,=\,\,\int
\,\,\frac{d \omega}{ 2 \pi \,i}\,\,
e^{\omega\,y \,\,+\,\,(\,\gamma( \oa )\,-\,1\,)\,r} \ti \p ( \omega )
\,\,=
\eeq
$$
\int \,\,\frac{d \omega}{ 2 \pi \,i}\,\,
e^{(\,\omega\,-\,\ol\,)\,y \,+\,(\,-\frac{1}{2} \,+\,\sqrt{\frac{
\oa\,\,-\,\,\ol}{\frac{\as N}{\pi} \,14 \zeta ( 3 )}}\,\,\,)\,\,r} \,\ti  \p (
\omega)\,\,.
$$
Evaluating the above integral using saddle point approximation we obtain
\beq \label{SADDLEOM}
\oa_{S} \,\,=\,\,\ol \,\,+\,\,\frac{1}{\frac{\as N}{\pi} \,14 \zeta ( 3
)}\cdot\frac{r^2}{4 \,y^2}\,\,
\eeq
which gives the answer:
\beq \label{SOLDIFF}
\p (y, k^2_t )\,\,=\,\,\frac{1}{\sqrt{k^2_t\,q^2_0}}\cdot\ti \p (\oa_{S})\cdot
\sqrt{\frac{2\,\pi\,(\,\oa_{S} -\ol\,)}{y}}\cdot e^{ \ol\,y\,-\,\frac{ln^2
\frac{k^2_t}{q^2_0}}{\frac{\as N}{\pi} \,28 \zeta ( 3
)\,y}}\,\,.
\eeq
We can trust this solution in the kinematic region where $ ( \,ln
\frac{k^2_t}{q^2_0} \,)^2\,\leq\,\frac{\as N}{\pi} \,28 \zeta ( 3
)\,y$. The solution of \eq{SOLDIFF} illustrates one very important property
of the BFKL equation, namely $k^2_t$ can be not only large, but with the
same probability it can  also be very small.
It means that if we started with sufficiently big value of virtuality $q^2_0$
at large value of $y\,=\,ln ( 1/x )$ due to evolution in $y$ the value of
$k^2_t$  could be small ( $k_t \,\sim \L$, where $\L$ is QCD scale ).
Therefore,
the BFKL equation is basically not perturbative  and the worse thing, is
that
we have not yet  learned what kind of assumption about the confinement
has been made in the BFKL equation.

Our strategy for the further presentation is to keep $k^2_t \,>\,q^2_0$ and to
study what  kind of nonperturbative effect we can expect on  including the
running $\as$ in the BFKL equation, as well as  changing  the value of
$\ol$ in the series of \eq{LIANDI}.

 \subsubsection{The bootstrap property of the BFKL equation.}
We have discussed the BFKL equation for the total cross section,
however this equation can also  be proved for the amplitude at transfer
momentum
$q^2 \,\neq\,0$, and not only for colorless state of two gluons in t-channel.
The general form of the BFKL equation  in
$\omega$ - representation  looks as follows \cite{BFKL} ( see \eq{MELLIN}:
 \beq \label{BFKLGEN}
(\, \omega\,\,-\,\,\omega^G ( k^2_t )\,\,-\,\,\omega^G ( ( q - k_t)^2 )\, )
\phi ( \omega, q, k_t )\,\,=\,\,\frac{ \as}{2  \pi} \,\l_R \int \frac{d^2
k'_t}{ \pi} K (q, k_t, k'_t ) \phi (\omega, q, k'_t)\,\,,
\eeq
where the kernel $K(q, k_t, k'_t )$ describes only gluon emission and
\beq \label{KERNELQ}
K ( q, k_t, k'_t )\,\,=\,\,\frac{k^2_t}{ ( k_t - k'_t )^2 \,k'^2_t }\,\,+\,\,
\frac{ ( q - k_t )^2}{ ( k_t - k'_t )^2 \,( q - k'_t )^2 }\,\,-\,\
\frac{q^2_t}{ ( q_t - k'_t )^2 \,k'^2_t }\,\,,
\eeq
$\l_R$ is colour factor where  $ \l_1 = 2 \l_8 = N $ for singlet and
$ ( N^2 - 1 )$ representations of colour SU(N) group and $\omega^G ( k^2_t
)\,\,=\,\, \a^G ( k^2_t )\,-\,1$ ( see \eq{REGGE} ).

The bootstrap equation means that the solution of the BFKL equation for
octet colour state of two gluons ( for colour SU(3) ), should give the
reggeizied gluon with the trajectory $\a^G (k^2_t)$ ( or $\omega^G (k^2_t )$
) given by \eq{REGGE}. The fact that the  gluon becomes a  Regge pole have
been
shown by us in the example of the next to the Born Approximation, and has
been used to get the BFKL equation in the singlet state. It means that the
solution of the BFKL equation in the octet state should have  the  form of a
Regge pole :
\beq \label{RP}
\p (\omega,q,k_t)\,\,=\,\,\frac{Const}{ \omega\,\,-\,\,\omega^G ( q^2 )}
\eeq
Assuming \eq{RP}, one arrives to the following  bootstrap equation:
\beq \label{BOOTSTRAP}
\, \omega^G ( q^2 )\,\,-\,\,\omega^G ( k^2_t )\,\,-\,\,\omega^G ( ( q -
k_t)^2 )\, \,\,=\,\,\frac{ \as}{2  \pi} \,\l_8 \int \frac{d^2
k'_t}{ \pi} K (q, k_t, k'_t ) \,\,,
\eeq
It is easy to check that the trajectory of \eq{REGGE} satisfies this equation.
It is interesting to mention that we can use \eq{BOOTSTRAP} to reconstruct
the form of the kernel $K (q, k_t, k'_t )$,  if we know the expression for
the trajectory ( see ref.\cite{BRA} ).
\subsection{Corect degrees of freedom at low $x$.}
In the previous subsections we gave the traditional derivation of the BFKL
 equation, which is rather difficult and demands a good experience
 in calculation of Feynman diagrams.
It is very important  to understand
 better the physical
meaning and the formal grounds of the BFKL equation. I firmly believe that
during this year A. Mueller \cite{MUEL} ( see also related papers
 \cite{NNN} ) has  achieved the considerable progress in
both understanding and formal derivation of the BFKL equation and its
generalization. Mueller's main idea is to construct the small $ x $
 infinite momentum  partonic wavefunction of a hadron in QCD while BFKL
calculated the amplitude for n - gluon production in so called multireggeon
kinematic region. The wavefunction  gives us much richer information on the
 hadron interaction, has very transparent physical meaning and makes
the bridge between our parton approach to ``hard" processes and  new phenomena
that we anticipate in the region of high density QCD.

{\bf 1.}

 The technical trick that has been used is also very instructive, namely it
  turns out that the wavefunction looks much simpler in the mixed
representation
in which each parton is labeled by its fraction of the total hadron momentum
$x_i$ and the transverse coordinate $r_{ti}$. The transverse coordinate
is especially useful since in the low $x_i$ region the $i^{th}$  gluon can be
considered as being emitted from the system of  $i - 1$ partons with
spatial transverse coordinates of these ``sources" being frozen during the
 emission  of $i- th $ gluon. Thus we can consider these  of $i - 1$
partons as a system of $ (i - 1 ) \,\, q \bar q $ dipoles since each gluon
can be  viewed as quark - antiquark pair if number of colours $ N_c$ is big
 enough. So the only thing that one  needs to write down is the emission of the
$i-th$ gluon
by such a system of dipoles. This problem has been  solved in
 ref.\cite{MUEL}.  For example for emission of gluon
  $ (x_2, r_{tG} = r_2 )$ from one dipole which is the
quark $ ( x_q = 1 - x_1, r_{t q} = r_0 =0)$  and antiquark $ (x_{\bar q} =
x_1, r_{t \bar q} = r_1)$ is equal to
\beq
\psi^{(1)} ( x_1, x_2 ;  r_1, r_2)\,\,=\,\,- \frac{i g T^a}{\pi}
\psi^{(0)} ( x_1 ; r_1 ) \lbrace \frac{r_{2\lambda}}{r^2_2}\,-\,
\frac{r_{21 \lambda}}{r^2_{21}} \rbrace \cdot \eps^{\lambda}_2\,\,.
\eeq

{\bf 2.}

Mueller made one very important step in our understanding of our parton system,
namely he found what sum rule plays the role of the momentum sum rules in the
GLAP approach for low $x_i$ partons. This sum rules is the normalization of
the partonic wavefunction:
\beq
\int \prod^n \frac{d x_i}{x_i} \prod^n d^2 r_{ti} |\Psi (x_1,...x_n; r_{t1},...
r_{tn}) |^2 \,\,=\,\,1\,\,.
\eeq

Using this equation one can easily
take into account so called virtual corrections,
 which in this case are mostly known as gluon reggeization or non -Sudakov
 form factor. The importance of this step can be compared only with the
transition from Gribov - Lipatov form of usual evolution equation with
Sudakov form factor in the kernel to well known  Lipatov - Altarelli -
 Parisi elegant form based on direct use of the momentum sum rules in QCD.

{\bf 3.}

The physical application of this new approach has not been considered but
Mueller noted at his Durham talk \cite{MUDURHAM}
 that his approach will be able to resolve
the old problem with the BFLK equation. Indeed, the physical meaning of
the growth of the structure function at $x_B \rightarrow 0 $ is the increase of
the number of ``wee" partons  ( N ) that can interact with the target ( $
N \propto x^{-\omega_0}_B $ )( see, for example, review \cite{LL}). However
the multiplicity of gluons calculated as the ratio $ \int \frac{ E d^3 \sigma}{
dA^3 p_{jet} }/\sigma_t $ turns out to be small ( of the order of $\alpha_s
\ln \frac{1}{x_B}$ ). It means that partons are in a very coherent state in a
typical inelastic event.   However the behaviour of the parton cascade at large
multiplicity should be quite different from  Poisson distribution since at
large multiplicity all N parton can be freed in the interaction.

{}.

The main progress was related to the fact that they discovered the correct
 degrees of freedom for  the region of small $x$ in QCD.
 They showed that if we discuss the deeply inelastic processes not in terms of
quark and gluons as we did before but introducing new degrees of freedom,
namely the color dipole of the definite size ($d$), we still can use the
 simple probabilistic interpretation in the region of low $x$. It means that
 the physical meaning of the deep inelastic structure function is not the
number of quark or gluons but the number of the color dipoles with the sizes
larger than $1/Q$ ($ d > \frac{1}{Q}$).
 It is worthwhile mentioning that at $x \sim 1 $ we still can use our old
 interpretation as probability to find parton  but as well as a new one.
$$\bullet$$
{\bf Problem 18:} Read Mueller's paper\cite{MUEL} and  try to understand
how simple looks the BFKL equation in new degrees of freedom.
$$\bullet$$

The main progress in the Mueller approach
 was related to the fact that he  discovered the correct
 degrees of freedom for  the region of small $x$ in QCD.
 He  showed that if we discuss the deeply inelastic processes not in terms of
quark and gluons as we did before but introducing new degrees of freedom,
namely the colour dipole of the definite size ($d$), we still can use the
 simple probabilistic interpretation in the region of low $x$. It means that
 the physical meaning of the deep inelastic structure function is not the
number of quark or gluons but the number of the colour dipoles with the sizes
larger than $1/Q$ ($ d > \frac{1}{Q}$).
 It is worthwhile mentioning that at $x \sim 1 $ we still can use our old
 interpretation as probability to find parton  but as well as a new one.

This idea shed a light on all technicalities of the BFKL equation which was the
manifestation of the artistic skill in the calculation of Feynman diagrams and
reduced the high art to normal physics understandable for a normal post-doc.
 I anticipate a big progress in the development of all kind problems related
to high energy QCD based on Mueller approach.

\section{New physical phenomena at small $x_B$:}
Let us outline the new phenomena which we anticipate
to occur in the region of small $x_B$ in perturbative QCD for
the case of deeply
inelastic scattering. Three of them are particularly important
and determine the physical picture of the parton evolution,
or cascade as we shall call it, in the region of small $x_B$, namely:
\par {\bf 1.}   The increase of the parton density \cite{BFKL}
 at $ x_B \rightarrow 0 $.
\par {\bf 2.} The growth of the mean transverse
momentum of a parton inside the parton cascade at low $x_B$
\cite{BFKL} \cite{GLR}.
\par {\bf 3.} The saturation of the parton density \cite{GLR}.

Before discussing each of these phenomena in turn,
let us try to understand these new phenomena at small $x_B$ by recalling
some facts regarding perturbation theory and deeply inelastic scattering.
According to the factorization theorem, the deep-inelastic
structure function is a product of a parton density and
a short distance coefficient function, which is
calculable in perturbation theory.
Both factors depend on the factorization scale, which we
choose equal to $Q^2$.
Concentrating on the gluon density, in the kinematical region
we are considering the former can be represented as
\beq \label{PERTSERIAS}
x G(x,Q^2) = \Sigma_n  C_n(Q_0^2) ( \alpha_s L )^n\,\, +\,\, O
( \alpha_s (\alpha_s L )^n)\,\,,
\eeq
where $L$ is the large logarithm in our problem. The coefficients
$C_n(Q_0^2)$ contain nonperturbative information and depend
on the initial scale $Q_0^2$ of the cascade, see Fig.28,

To see what kind of large logarithm could occur, one can
examine the
probability of emission $P_i$ of the i-th parton in the cascade.
Near $x = 0$ it can be written as
$$ \label{emission}
P_i = \frac {N_c\alpha_s}{\pi } \cdot
 \frac{dx_i}{x_i} \cdot \frac{d q^2_{it}}{q^2_{it}}$$
\par
1. In the region of large virtualities of the photon
in deep inelastic scattering (but not at very small values of $x_B$)
$L$ is equal to $\ln Q^2$.
\par
2. At $Q^2$ fixed  ($ Q^2 \,\,\simeq\,\,Q^2_0\,\,\gg \,m_{proton}^2$)
but at small value of $x_B$
we have a different large logarithm in eq.(2), namely
$ L \,=\, \ln \frac{1}{x_B}$.
\par
3. When both $ \ln Q^2$   and $\ln \frac{1}{x_B}$ are
large ($Q^2 \gg Q^2_0\,\, x_B  \rightarrow 0 $) the large
logarithm
has a more complicated form, namely $L \,=\,\ln Q^2 \ln \frac{1}{x_B}
$. In this case one can apply the so-called double logarithmic
approximation (DLA)
of pQCD. The objective now is to resum these large logarithms
in eq.(9), so that we can estimate the behavior of the parton densities
in the corresponding kinematical regimes.
This resummation is performed via evolution equations,
which differ depending on the type of large logarithm under
consideration.
 \par   \subsection
{Increase of the gluon (quark) density at $x_B \rightarrow 0$.}
We now investigate the new phenomena at small $x_B$, but
begin by discussing the evolution equations that sum
particular classes of large logarithms in the parton densities.

\par
 {\em Gribov - Lipatov - Altarelli - Parisi (GLAP) evolution equation.}
\par The GLAP equation sums those contributions in the parton
cascade which compensate the smallness of the QCD coupling
constant $\alpha_s$ by the large logarithm $\ln Q^2$.
It reads
\beq
\frac{d}{d\ln Q^2}\p(x,Q^2)=
\a_s(Q^2)\int_x^1\frac{dz}{z}P(z)\p(\frac{x}{z},Q^2)
\eeq
where $P(z)$ is the Altarelli-Parisi evolution kernel, calculable
in perturbation theory, and $\phi(x,Q^2)$ is a generic
parton density (matrix).

Clearly, from the expression for $P_i$,
we need strong ordering ($Q^2 \gg ...\gg q^2_{it}\gg .....
\gg \frac{1}{R^2_p} = Q^2_0$)
in transverse momenta of emitted
partons since only under such a condition does one get a
$\ln Q^2$ contribution for each integration over $q_{it}$ in
the parton cascade in Fig.28.
This strong ordering means that the GLAP evolution equation allows us to
calculate the probability to find a parton of transverse size
$r_t \approx \frac{1}{Q} $ inside the initial parton (quark or gluon)
in the hadron at fixed $x_B$ (see Fig.29).
In the case where both $Q^2$ and $1/x_B$ are very large (the DLA
approximation), the AP kernel simplifies to $P(z)= N_c\alpha_s/\pi z$,
and the solution to the GLAP equation for the gluon density reads then
$$
xG(x,Q^2) \sim x G(\omega_0,Q_0^2) \,
\exp(\sqrt{\frac{4N_c\alpha_s}{\pi}\ln\frac{Q^2}{Q_0^2}\ln\frac{1}{x}})
\sqrt{\frac{\omega_0^3}{3 \alpha_s \ln\frac{Q^2}{Q_0^2}}},
$$
where $\omega_0=(\frac{N_c \alpha_s}{\pi}
\ln\frac{Q^2}{Q_0^2}/\ln\frac{1}{x})^{1/2}$,
and $Q_0$ is the initial scale of the parton cascade.
$$\bullet$$
{\bf Problem 19:} Find the exact solution of the GLAP equation at low $x$ (
see eq, (94)), using $P(z)|_{z \ra 0}\,\ra \,\frac{N_c \as}{\pi z}$.
$$\bullet$$

Note that the solution of the GLAP
equation decreases as $Q^2$ increases. This
agrees with the intuitive notion that, because of
asymptotic freedom, the
hadron should be almost empty at small distances.
\par
{\em Balitski- Fadin -Kuraev - Lipatov (BFKL) evolution.}
\par
The situation changes crucially if we consider the behavior
of the parton density $\phi$ in another extreme regime:  at
fixed $Q^2$ but small $x_B$. Here we have the
large logarithm $L\,\,=\,\,\ln\frac{1}{x_B}$.
The picture of the parton cascade, Fig.30, shows that the number
of partons increases drastically in the region of small $x_B$,
since each parton in the basic branching process in Fig.28
is allowed, due to the abundance of available energy,
to decay in its own chain of daughter partons.
Let us estimate the total multiplicity of gluons $N_G$ associated
with this complicated process.

We are able to
estimate the number of cells in this chain diagram,
which corresponds to the typical number of parton (gluon) emissions in
one {\it sub}process in our parton cascade  ($\overline{n_G}$).
The characteristic value of $\Delta y_{i,i+1} $,
the rapidity difference between two adjacent rungs of the ladder,
is equal to
$ \Delta y_{i,i+1}\,\,\propto 1/\alpha_s\,\,,$
from the expression for gluon emission $P_i$.
Thus
\beq
\overline{n_G}
\,\,=\,\,\frac{\ln \frac{1}{x_B}}{\Delta y_{i,i+1}}\,\,\propto\,\,
\alpha_s \cdot \ln \frac{1}{x_B}\,\,.
\eeq
The total number of $N_G$ partons that could interact with the target
can then be calculated as follows (see Fig.30):
\beq
N_G\,\,\propto \,\,e^{\overline{n_G}} \,\,=\,\,e^{c\alpha_s\ln 1/x_B }\,\,
=\,\,(\frac{1}{x_B})^{c\alpha_s}\,\,,
\eeq
where the constant $c$ should be calculated using the exact
BFKL equation \cite{BFKL}. In first approximation $c\alpha_s=0.5$.

\subsection
{A new scale for the deeply inelastic process at $x_B
\rightarrow 0$.}

Thus in the region of small $x_B$
the density of partons increases.
A careful  study of the behavior of the parton cascade shows
that also the mean transverse momentum of
the parton increases.
The reason is as follows.
In the case where we neglect the running coupling constant of QCD
our theory is dimensionless so each emission leads to a value of the
transverse momenta of the daughter gluons which is of the same order as the
transverse momentum of the parent gluon. This  could be seen just
from the explicit expression of $P_i$. We introduce the ratio
$ |\ln {q^2_{i,t}}/{q^2_{i+1,t}}|$ which characterizes the emission.
It is roughly constant.

After $n_G$ emissions
\beq
<|\ln (q^2_{n,t}/ Q^2_0)|> \,\,\propto \,\,\sqrt n_G \,\,
 \approx \sqrt{\alpha_s \ln \frac{1}{x_B}}\,\,,
\eeq
since the parton cascade corresponds to a random walk in the variable
$\ln ( q^2_{n,t}/Q^2_0)$ \cite{BFKL}\cite{GLRNUCL}
(see Fig.30).  Although we will not explicitly
present the BFKL equation here, we will give the solution to this
equation with the initial condition $x_B G(x_B=x_0) = \delta(\ln(Q^2/Q_0^2))$.
This is instructive because it shows explicitly all
the properties mentioned in the above:
$$
G(y-y_0,r-r_0)=\sqrt{\frac{Q^2}{Q_0^2}}\cdot
\frac{1}{\sqrt{\pi\Delta\omega(y-y_0)}}\cdot
\exp\{\omega_0(y-y_0)-\frac{(r-r_0)^2}{8\Delta
\omega_0(y-y_0)}\}
$$
where
$$
\omega_0=\frac{\alpha_s(Q_0^2)N_c}{\pi}4\ln 2, \;
\Delta=\frac{14\zeta(3)}{4\ln 2},\; y-y_0=\ln\frac{x_0}{x_B},
\; r-r_0=\ln\frac{Q^2}{Q_0^2}
$$
$$\bullet$$
{\bf Problem 20:} Calculate the mean value of $\ln (Q^2/Q^2_0)$ using the
 solution of the BFKL equation.
$$\bullet$$
\subsection{Saturation of the gluon density.}

The increase of the parton density leads to a new problem in
deeply inelastic scattering, namely the violation of
s-channel unitarity.
This is the requirement that the total cross section for virtual
photon absorption be smaller than the size of a hadron.
\beq
\sigma (\gamma^* N) \ll \pi R^2_h
\eeq
At small $x_B$ the gluon density
dominates and we can write
\beq
\sigma (\gamma^* N) \, \, =\, \, const\frac{\alpha_{em}}{Q^2}
 \,\,   x G(x, Q^2)
\eeq
where $const\,\,\alpha_{em}/Q^2$ approximates the cross section
$\sigma(\gamma^* g)$.
We have shown previously that the value of the gluon density
increases
very rapidly as $x \rightarrow 0$. Using the DLA result (see eq.(11)),
we can rewrite the
unitarity constraint in the following way, dropping the $const$:
\beq \label{UNITARITY}
\frac{\alpha_s(Q^2)}{Q^2}\cdot
(\frac{1}{x_B})^{c \alpha_s(Q^2)}
\leq \pi R^2_N.
\eeq
Here we replaced $\alpha_{em}$ by $\alpha_s$ since the probe can
also be a virtual gluon, not just a virtual photon.

{}From this expression alone one can conclude that {\bf unitarity
will be violated}  \cite{GLR}  at
\beq
x < x_{{\rm cr}} \,\,\,\,{\rm where}\,\,\,\,
\log \frac{1}{x_{{\rm cr}}}\, \, =\, \, \frac{1}{c} \cdot
       \frac{1}{\alpha_s(Q^2)} \ln(\frac{\pi R_N^2 Q^2}{\alpha_s(Q^2)})
\eeq
Therefore  unitarity is violated even for (very) large
values of $Q^2$ when $x < x_{{\rm cr}}$.
Clearly the miraculous confinement force
cannot prevent this from happening. Thus we have to look for
the origin and solution of this problem within pQCD.

Let us try to understand what happens in the region of small $x_B$ by
examining the parton distribution in the transverse plane (see Fig.29)
\footnote{Recall that a high energy hadron
in the parton model can be represented as a Lorentz contracted disc.}.
Our probe (photon) feels those partons whose size is of the order
of $\frac{1}{Q}$.
To begin, at $x \sim 1$ we have only several partons that are distributed in
the hadronic disc.
If we choose $Q^2$ such that
\beq
  r^2_p \approx \frac{1}{Q^2} \, \, \, \ll \, \, \,  R^2_h.
\eeq
then the distance between partons in the transverse plane is much larger
than their  size and we can neglect the interaction between
partons. The only process which is essential here is the emission
of partons that is  taken into account in the usual evolution
equation.
As $x$ decreases the number of partons increases and at some value
of $x = x_{{\rm cr}}$ partons start to densely populate the whole
hadron disc.
For $x < x_{{\rm cr}}$ the
partons overlap spatially and begin to interact
throughout the disc. For such small $x$-values
the processes of
recombination and annihilation of partons should be as essential as
their emission. Both these processes are however not
incorporated into either the GLAP or  BFKL evolution equations.

What happens in the kinematical region $x < x_{cr}$
is anybody's guess
but there is enough experience with some models \cite{MOD} to
suggest the
so-called {\bf saturation of the parton density} \cite{GLR}.
This means that the parton density $\phi$,
which we discussed before, is constant in this domain.

\section{Shadowing correction and nonlinear evolution equation.}
\subsection{Qualitative Derivation of Nonlinear Gribov-Levin-Ryskin
Evolution Equation}
In section 11.3 we remarked that at small $x$  (high
energies),
the density of gluons becomes so large that the unitarity constraint is
violated, even at large values of $Q^2$.
We argued  that the physical  processes
of interaction and recombination of partons, which
are usually omitted in perturbative calculations, become important
in the parton cascade at a large value of the parton density.
To take interaction and recombination into account we must identify a new
small parameter that lets us estimate the accuracy of our calculations.
This  small parameter is \cite{GLR}
\beq \label{W}
 W\,\,=\,\,\frac{\alpha_s}{Q^2}\,\,  \rho(x,Q^2)\,\,.
\eeq
The first factor
is the cross section for absorption of a gluon by a
parton in the hadron; the transverse density $\rho$ is defined
in  eq.( \ref{DEFINITION1}).  Effectively, $W$
is the probability of a parton recombination during
the cascade. We can rewrite the unitarity constraint
 eq.(\ref{UNITARITY}) in the form
\beq
 W\,\,\leq\,\,1\,\,.
\eeq
Thus, $ W $ is the natural small parameter in our problem.
Amplitudes that take gluon recombination into account
can be expressed as a perturbation series in this parameter \cite{GLR}.
We can resum this series,
and the result
\cite{GLR} can be understood easily
by considering the structure of the QCD cascade in a fast hadron.
Two processes occur inside the cascade
\beq
 {\rm emission} \,\,\,\,(\, 1\,\rightarrow \,2\,);\,\,\, \,\,{\rm probability}
\,\,\propto\,\,\a_s \,\rho\,\,\,;
 \eeq
$$
{\rm annihilation}\,\,\,\,(\, 2\,\rightarrow\,1\,);\,\,\,
 \,\,{\rm probability}\,\,\propto\,\,\a^2_s\,
r^2\, \rho^2 \,\,\propto\,\,\a_s^2\,\frac{1}{Q^2}\, \rho^2\,\,,
$$
where $r$ is the size of a parton produced in the annihilation process. For
deep inelastic scattering $ r^2 \propto 1/Q^2$.
\par
At $x\, \sim \, 1 $ only the production of new partons
(emission) is essential because $ \rho \,\ll\,1$, but at
$x\,\rightarrow\,0$
the value of $\rho$ becomes so large that the annihilation
of partons becomes important.
This simple parton picture allows us to write
an equation for the density of partons that properly accounts
for all these processes.
The number of partons in a phase space cell
($\D y = \D \ln(1/x), \D \ln Q^2 $) increases through
emission and decreases through annihilation. Thus the
balance equation reads
\beq
\frac{\partial^2 \rho}{\partial \ln\frac{1}{x}\,\partial \ln Q^2}\,\,=
\frac{\alpha_s N_c}{\pi} \,\rho\,\,-\,\,\frac{\alpha^2_s
\,\tilde{\gamma}}{Q^2}\,\rho^2\,\,,
\eeq
or in terms of the gluon structure function $x G ( x, Q^2 ) $
\beq \label{GLR}
\frac{\partial^2 x G ( x,Q^2 )}{\partial \ln \frac{1}{x}\,\partial
\ln Q^2}\,\,=\,\,\frac{\alpha_s N_c}{\pi} x G (x, Q^2 )\,\,-\,\,
\frac{\alpha^2_s \,\tilde{\gamma}}{\pi Q^2 \,R^2} \,( x G (x,Q^2 ) )^2\,\,.
\eeq
This is the so-called Gribov-Levin-Ryskin (GLR) equation \cite{GLR}.
The parameter $\tilde{\gamma}$ can in fact be calculated order by order
in $W$ perturbation theory. Mueller \& Qiu \cite{MUQI} found it to be
$$
\tilde{\gamma}\,\,\,=\,\,\,\frac{81}{16} \,\,\,\,\,{\rm for} \,\,N_c\,\,=3.
$$
\subsection {The Scale of the Shadowing Corrections}
The second term in Equation \ref{GLR} describes shadowing
corrections (SC). Its size  crucially depends
on the value of $R^2$. The physical meaning of $R^2$ is clear:
$R$ is the correlation length between two
gluons in a typical hadronic situation (at $x \sim 1 $).
In our derivation of the GLR equation we assumed
that there are no correlations between gluons except those imposed
by their confinement in a disc of radius $R$.
\par  If $R \,=\,R_h$
the value of the SC is negligibly small.
\par  If $ R\,\ll\,R_h$,
the SC could be large (see Refs. \cite{LOXR}, \cite{KMRS} and \cite{EXNE}).
Recently Braun et al \cite{BRAUN} performed the
first theoretical estimates of the value
of $R$ within the framework of QCD sum
rules. They found
$$
R\,\,=\,\,0.3 - 0.35 \,fm \,\,\sim\,\,\frac{1}{3}\, R_h\,\,.
$$
an encouraging result for experimental study of SC at present energies.
 For the gluon structure function the solution of the GLR equation
is shown in Fig.32 and the value of $x G (x,Q^2)$ from which saturation
starts depends on the value of $R^2$ . For $R = R_p$  saturation starts
from
$$ xG (x,Q^2) > 150 \,\,\,\,at\,\,\,\, Q^2 = 10 GeV^{-2}$$
while for $R = \frac{1}{3} R_p $
$$ xG (x,Q^2) > 15 \,\,\,\,at\,\,\,\,Q^2 = 10 GeV^{-2}\,\,.$$
It should be stressed that the GLR equation predicts the value of the maximal
packing factor ( PF) which is
\beq
PF\,\,\,=\,\,< | r^2_{constituent}|> \cdot \rho
\eeq
PF does not depend on the value of the radius $R$ and for parton with $
<|r^2_{constituent}|> \,\,=\,\,\frac{1}{Q^2}$ the GLR equation gives:
$$
( PF )_{max}\,\,=\,\,0.21\,\,\,\,for\,\,\,\,N_c\,\,=\,\,3$$

\subsection{New Scale of Transverse Momentum (the Critical Line)}
\subsubsection{The main property of a solution of the GLR equation}
First, let us rewrite the GLR equation (see eq.( \ref{GLR})) in terms of
new variables: $ F = xG(x,Q^2)$; $y\,=\,8N_c/b\cdot\ln(1/x)$ and
$\xi \,=\,\ln \ln (Q^2/\L^2)$, as
\beq \label{GLRNEWVAR}
\frac{\pa^2 F}{\pa y\pa \xi}\,\,=\,\,\frac{F}{2}\cdot \{ \,1\,\,-\,\,\gamma'
 exp [\, - \xi \,-\,e^{\xi}\,]\,F\}\,\,,
\eeq
where  $\gamma'\,\,=\,\,2\pi \tilde{\gamma}/b R^2 \L^2$.
The terms $e^{\xi}$
( $ - \xi$)  in the exponential correspond $1/Q^2$ ($\a_s$).
Note that the term proportional to $\gamma'$ corresponds
to the recombination probability $W$, defined below Equation (26).
By inspecting eq.(\ref{GLRNEWVAR}) we can guess the main property
of the solution. At the start of the evolution, when the value of the
structure function is small, we can neglect the nonlinear term. The
derivative with respect to $y$ is here positive, so the structure function
increases monotonically with $y$, or $\ln(1/x)$. This growth
continues up to the value
$$F_{max}\,\,=\,\,\frac{1}{\gamma'} \cdot exp [ \xi\,+\,e^{\xi}\,],$$
where the derivative vanishes. If $F > F_{max}$ the derivative
from the GLR equation becomes negative and $F_{max}$ is thus the limiting value
of the gluon density and the solution of the GLR equation as $x \ra 0$.

Therefore, at each value of $Q^2$ there is a critical value of
$x \equiv x_{cr}$
when $F = F_{max}$ or, conversely, at each value of $x$ there is a
critical value of $Q^2 \,=\,q^2_0 (x ) $ when $ F = F_{max}$. This value
$Q^2 = q^2_0 (x)$ introduces a new transverse momentum scale in the parton
cascade. The behavior of the gluon density changes
crucially  at $Q^2 = O(q^2_0 (x))$. Indeed,
$$F = F^{GLAP} (x, \ln Q^2 ) \,\,\,\,\,\,\,{\rm for}
\;\;  Q^2\, >\, q^2_0 (x) ,$$
where $F^{GLAP}$ means that $F$ is a solution of the linear GLAP equation
and a smooth function of $\ln Q^2$. Alternatively
$$ F \,\propto\,\,\frac{ Q^2}{q^2_0(x)} \,\,\,\,\,\,{\rm for}
\;\; Q^2 \,<\, q^2_0 (x) F_{max}\,.$$
We postpone the discussion of the physical meaning of this new scale
and instead determine first the $x$ dependence of $q^2_0 (x)$.
\subsubsection{Trajectories}
A semiclassical analysis offers new insights into solutions
of the GLR equation.
It is easiest
to find the solution to the linear equation by passing to a
moment ($\omega$) representation
and
introducing the anomalous dimension $\gamma$.
In moment representation the solution is as follows:
\beq \label{LINTRA}
F (x,Q^2) \,\,=\,\,\int \frac{d \omega}{2\pi i}
M_0 (\omega,Q^2 ) e^{\omega y \,\,+\,\,\gamma(\omega)\xi}
\,\,,
\eeq
where $M_0 (\omega,Q^2)$
is the initial distribution at $Q^2 = Q^2_0 $
and the anomalous dimension
$\gamma (\omega) \,=\,1/2\omega$. At large $y$ (small $x$) we
can evaluate the integral in the saddle point approximation. The position
of the saddle point is determined by:
\beq \label{LINTRAJECTORY}
y \,\,=\,\,-\,\frac{d \ln M_0 (\omega)}{d \omega} \,\,-\,\,
\frac{d\gamma(\omega)}{d \omega} \cdot \xi\,\,\,.
\eeq
This equation describes a family of curves
$y \,=\,y (\xi )$ that can be considered as semiclassical trajectories of
parton evolution because  $\omega$ is a constant of
a motion in the
linear evolution equation. The semiclassical values of $F$
along the trajectory give an approximate estimate of the structure function at
fixed $Q^2$ and  $x$.
Although this method yields no new understanding
for the linear equation,
it can be generalized to the nonlinear case.
\subsubsection{Semiclassical Solution to the GLR Equation}
In semiclassical approximation we parametrize the solution to the
GLR equation by writing
$$F\,\,\,=\,\,e^S, \,\,\,\,\, {\rm assuming}\,\,\,
S_y\, S_{\xi} \,\,\gg\,\,S_{y \xi}
\,\,,$$
where $S_y \,=\,\pa S/\pa y$ ,  $S_{\xi} \,=\,\pa S/\pa \xi$
and $S_{y \xi}\,=\,\pa^2 S/\pa y \pa \xi$.
The solution to the linear equation is $S = \sqrt{2 y \xi}$, for which
one can check that the above
inequality holds to good accuracy.

Thus in the semiclassical case the GLR equation can be reduced to the form:
\beq \label{SEMICLASEQ}
S_y \, S_{\xi} \,\,=\,\,\frac{1}{2}(1 \,\,-\,\,\gamma' \,\,exp\{ \,S\,\,-\,\,
e^{\xi}\,\,-\,\,\xi\,\}) \,\,\,.
\eeq
Eq.( \ref{SEMICLASEQ}) can be solved using the method of characteristics,
or trajectories (\cite{GLR} \cite{BARBUSHU} \cite{COLKWE}).
By introducing an intrinsic
time coordinate $t$ along the trajectory and denoting the derivatives with
respect to this time variable
by $\dot{y}$ etc we can rewrite the GLR equation as the following system of
ordinary equations:
\bea \label{SEMICLASSSYSTEM}
\dot{y}\,\,=\,\,S_{\xi} \nonumber\\
\dot{\xi}\,\,=\,\,S_y\nonumber\\
\dot{S}\,\,=\,\,2 \,S_y\,S_{\xi}\nonumber\\
\dot{S_y}\,\,=\,\,-\,\gamma'\,\,S_y
\,\,exp\{\,S\,\,-\,\,e^{\xi}\,\,-\,\,\xi\,\}
\nonumber\\
\dot{S_{\xi}}\,\,=\,\,-\,\gamma'\,\,[\,S_{\xi}\,-\,1\,\,-\,\,e^{\xi}\,]
\,\,exp\{\,S\,\,-\,\,e^{\xi}\,\,-\,\,\xi\,\}\,\,.
\eea
In the linear case ($\gamma'=0$)
this system of equations has a simple solution:
$$ S_y \,=\,\omega\,=\,Const\,\,;\,\,\,\,S_{\xi}\,\,=\,\,\frac{1}{2\omega}\,=
\,Const\,\,,$$
which coincides with eq.(\ref{LINTRAJECTORY}).
We introduce $\omega$ here such that it corresponds to
$\omega$ in eq. (111). From eq.(113 ) we find
$$
y-y(t=0) = \frac{1}{2\omega}t \;\;,\;\;
\xi-\xi(t=0) = \omega t \;\;,\;\;
S = \frac{1}{2}t
$$
Thus we obtain
$$
S = \sqrt{2(y-y(t=0))(\xi-\xi(t=0))} \;\;,\;\;
y-y(t=0) = \frac{1}{2\omega^2}(\xi-\xi(t=0))
$$
The second of these equations corresponds to Equation 33
with $\gamma = 1/\omega$, the correct anomalous dimension
in the GLAP equation for $x\rightarrow 0$.
Along each trajectory of the
linear equation, the rescattering probability $W \,=\,\gamma' \,exp\{S \,-\,
e^{\xi}\,-\,\xi\,\}$ increases at the beginning when $\dot{S} \,\approx\,1$,
reaches a maximum value when $S_{\xi} \,=\,( 1\,+\,e^{\xi}\,)$ and then
decreases at larger $\xi$. However, the trajectories of the GLR equation become
quite different from the linear ones. Indeed, in the region where $W$ becomes
of order one, the derivatives $S_y$ and $S_{\xi}$ are not small and
trajectories are not straight lines. A number of
papers (\cite{GLR} \cite{BARBUSHU} \cite{COLKWE}) have shown that
critical line separates
the region of almost-linear trajectories (small values of $W  \leq 0 (\a_s)$)
and strongly nonlinear solutions (see Fig.33).

It is easy to check explicitly that for large $\xi$,
\beq \label{CRITLINE}
y_{cr}\,=\,\frac{e^{2\xi}}{4}\,;\,\,\,S\,=\,e^{\xi}\,\,-\,\,\ln \gamma'\,;
\,\,\,S_y\,=\,e^{-\xi}\,;\,\,S_{\xi}\,\approx\,\frac{e^{\xi}}{2}\,\,
\eeq
is the solution of eq.(\ref{SEMICLASSSYSTEM}) and of
the GLR equation as well.
To the right of the critical line
($x > x_{cr}$)  the equation $y_{cr} = \frac{e^{2\xi}}{4}$
may be recast as
\beq \label{CRITICALLINE}
\ln \frac{1}{x_{cr}}\,\,=\,\,\frac{b}{32 N_c} \ln^2( Q^2/\L^2 ).
\eeq
The nonlinear corrections are so small that we can neglect them and
use the trajectory of the standard
GLAP equation. However, when $x$ becomes
smaller than $x_{cr}$, $W \rightarrow 1$ and we have to use the nonlinear
trajectory. Remarkably, this trajectory cannot cross the critical
trajectory (see Fig.33). Using this property, one can suggest the
following way to solve the GLR equation.
To the right of the critical line it suffices to find the solution of
the linear (GLAP) equation with the new boundary condition $ F\, = \,e^{S} $
with $S\,=\,e^{\xi} - ln \gamma'$ on the critical line
(see eq.(\ref{CRITLINE})).
To the left of the critical line we have a separate system of trajectories.
The solutions to the left do not depend on the solutions to
the right of the critical line. Here we enter the very
interesting kinematic region of hdQCD. Even the GLR equation
is not the right tool to solve problems in this region
because in addition to the recombination
of two partons, simultaneous interactions
of three, four and more partons are also important in this region.

\subsection{The Physical Meaning of the New Scale for the Typical Transverse
Momentum in the Parton Cascade}
The equation for the critical line
(eq.(\ref{CRITICALLINE})) introduces a new
scale for the value of the typical transverse momentum in the parton cascade,
namely
\beq
 q^2_t\,\,=\,\,q^2_0 (x) \, \v_{x\,\ra\,0}\,\,\ra
\,\,\L^2\,\,e^{\sqrt{\frac{32N_c}
{b} \,\ln \frac{1}{x}}}\,\,.
\eeq
This new scale plays the role of an
infrared cutoff in inclusive production.
This new infrared cutoff has a dynamical
origin:  parton-parton recombination in the parton cascade. It is not
related to confinement.

The quantity $q_0 (x)$ is also
the Landau-Pomeranchuk momentum \cite{LP} in the parton medium. Indeed, in  the
saturation region $q_t < q_0 (x)$  the mean free path $\lambda$
does not depend on $x$.   In this region
$\lambda \,=\,1/ \s T $ where $\s $ is the cross-section of parton-parton
 interaction ($\s \propto \a_s 1/q^2_t$) and $ T = xG (x,q_t)$
 is the number of
 partons with transverse momentum $q_t$  at fixed $x$.
 If $q_t < q_0 (x) $  $T \propto q^2_t \p_0$ and
$\s T = Const (q_t)$.
Thus $\lambda$ is equal to the formation length of a parton
with  $x = x_0 $, which can be derived from the relation $ q_0 (x_0) =q_t$.
If $q_t > q_0 (x)$  then  $\s T \,\propto x G(x,q_t )/q^2_t  \ll 1/mx $
so $\lambda $ is big. Therefore, $ q_0 (x) $ is precisely the LP momentum, and
the factor C that interpolates between coherent and incoherent
emission of gluon with transverse
momentum $q_t$ and fraction of energy $x_j$
can be rewritten as follows:
\beq
C\,\,=\,\, \frac{1}{ 1\,\,\,+\,\,\,\frac{ x_0 (q_t)}{ x } \cdot
 \frac{q_0 (x_j)}{q_t} }\,\,.
\eeq

\subsection {Correlations}

We assumed that the only correlations between gluons
in the cascade arise from
the fact that the gluons are distributed in a disc of radius $R$.
The probability to find two partons with the same value of
$\ln(1/x)$  and $ \ln Q^2$ is
\beq
P_2\,\,\propto \,\,\rho^2\,\,.\label{rhosq}
\eeq
However this assumption is only correct if the
number of colors is large ($N_c \rightarrow \infty$) \cite{TWFOUR}.
If $N_c$ is not large, Bartels and Levin et al claim  \cite{TWFOUR} claim that
\beq \label{TWOCORRELATIONS}
\frac{ P_2}{\rho^2} \,\,\rightarrow \,\,\exp\, (\sqrt{\,\,
\frac{4 N_c \alpha_s}{\pi}
\,\,\ln \frac{1}{x}\,\ln
\frac{Q^2}{Q^2_0}\,\,}\frac{1}{(\,N^2_c\,-\,1)^2 })\,\,.
\eeq
Bartels \& Ryskin \cite{BARYS} found that this correlation
reduces the correlation radius $R$ by a factor of
1.3 - 1.4 in the HERA
kinematic region.

It is also very instructive to write down the evolution equation that takes
into  account the correlation of eq.( \ref{TWOCORRELATIONS}).
This is a system of two equations, instead of the single equation
of eq. ( \ref{GLR}), namely
\bea
\frac{\partial^2 \rho}{\partial \ln\frac{1}{x}\,\partial \ln Q^2}\,\, & = &
\frac{\alpha_s N_c}{\pi} \,\rho\,\,-\,\,\frac{\alpha^2_s
\,\tilde{\gamma}}{Q^2}\,P_2\,\,, \\
\frac{\partial^2 P_2}{\partial \ln\frac{1}{x}\,\partial \ln Q^2}\,\, & = &
\frac{4\alpha_s N_c}{\pi}(1+\frac{1}{(N_c^2-1)}) \,P_2\,\,-
\,\,\frac{2\alpha^2_s
\,\tilde{\gamma}}{Q^2}\,P_2 \rho\,\,.\label{43}
\eea
Here we assume that the probability to find three
partons in the same cell of the parton phase space
is $P_3 = P_2\cdot \rho$.
Without this assumption the last term in Equation (\ref{43}) reads
as $2\alpha_s^2 \gamma/Q^2 \cdot P_3$. The general equation
for $P_n$ is (see Ref.\cite{EXNE} for details)
\beq
\frac{\partial^2 P_n}{\partial \ln\frac{1}{x}\,\partial \ln Q^2}\,\, =
\omega \gamma_n P_n - \frac{n \alpha_s^2 \gamma}{Q^2}P_{n+1}
\label{Pneqn}
\eeq
This equation indicates the need for the solution of two theoretical
problems.

(i) We should find a general equation for $\gamma_n$, which
has the formal meaning of the anomalous dimension of a
high twist ($2n$) operator (Refs.\cite{EXNE},\cite{HIGHTW},
\cite{TWFOUR}).

(ii) We need a closed form of the equation which incorporates
all $P_n$ for the deep-inelastic structure function, and
its solution.

The first problem was solved by A.G. Shuvaev and us in
Ref.\cite{HIGHTW}. The complicated problem of gluon-gluon
interaction was reduced to that of interaction of colorless
``gluon ladders'' (Pomerons). It was shown in Ref.\cite{TWFOUR}
that this idea works for the case of the anomalous dimension
of the twist four operator, in other words, for the case of
$P_2$. The fact that we can consider the rescattering of
$n$ Pomerons to find $\gamma_n$ (or $P_n$) signifies
that we are dealing with a quantum problem:
the calculation of the ground state energy for an
$n$-particle bosonic system where the interaction is attractive
and given by a four-Pomeron contact term, with a coupling
proportional to $\lambda=4\alpha_s/(N_c^2-1)$.
This observation simplifies the problem and enabled us
to reduce it to solving the nonlinear Schr\"{o}dinger equation
for $n$ Pomerons in the t-channel. It should be stressed
that this effective theory is a two-dimensional one,
the two dimensions being spanned by $\ln(1/x)$
and $\ln Q^2$:
the Schr\"{o}dinger equation can be written for $n$ particles
on a line. It is well-known that this problem can
be solved exactly (e.g. by Bethe Ansatz) and the answer was found to be
\beq
\gamma_n = \frac{N_c \alpha_s n^2}{\omega \pi}[1+
\frac{1}{3}\frac{1}{(N_c^2-1)^2}(n^2-1)]
\eeq
Although Pomerons are bosonic, maximally two Pomerons occupy a single-particle
level in this groundstate configuration, thus behaving like
fermions. This is due to the two-dimensional nature of
this problem.
The second problem has also been (partly) solved by us (unpublished
but see Ref.\cite{EXNE} for results). We introduce the
generating function
\beq
P(x,Q^2,\eta) = \sum_{n=1}^{\infty} P_n e^{n\eta}
\eeq
with $P_1=\rho$. Using eq.(\ref{Pneqn}) we obtain
\bea
\frac{\partial^2 P}{\partial \ln\frac{1}{x}\,\partial \ln Q^2}\,\, & = &
\frac{N_c\alpha_s}{\pi}(P_{\eta\eta} + \frac{1}{3}\frac{1}{(N_c^2-1)^2}
(P_{\eta\eta\eta\eta}-P_{\eta\eta}) \nonumber \\
& & - \alpha_s^2\gamma e^{-\ln(Q^2)}e^{-\eta}(P_{\eta}-P), \label{newevol}
\eea
where $P_{\eta}=\partial P/\partial \eta$, $P_{\eta\eta}
=\partial^2 P/\partial \eta^2$ etc. The deep-inelastic
structure function is the solution of this equation
at $\eta=-\ln(Q^2)$, viz.
\beq
x\, G(x,Q^2) = \pi R^2 P(x,Q^2,\eta=-\ln(Q^2).
\eeq
Eq.(\ref{newevol}) is a new evolution equation which allows
one to take into account both initial and dynamical correlations
between gluons in the parton cascade. The key problem in
solving it is nonperturbative in nature, namely to determine
an initial condition at $x = x^0$. The dependence of
$P(x^0,Q^2,\eta)$ on $\eta$ describes the correlations between
gluons in the hadron near $x \sim 1$. The simplest case is
to neglect this initial correlation and use the eikonal
approach for $P(x^0,Q^2,\eta)$, namely
\beq
P(x^0,Q^2,\eta) = \sum_{n=1}^{\infty} e^{n\eta} \frac{(-1)^{n+1}}{n!}
P_1(x^0,Q^2)^n
= 1-\exp(-e^{\eta}P_1(x^0,Q^2))
\eeq
The solution of this equation has not yet been found (see Ref.\cite{EXNE}
for more details).

\section{A look at the first small $x$ HERA data.}
\subsection{25 $nb^{-1}$ of Experimental Data}
Let us list conclusions from the first experimental results at small $x$
from the H1 and ZEUS collaborations at HERA (\cite{HERA}):
\par
1. The increase of the value of the deep inelastic structure function $F_2$
at $x\,\rightarrow \,10^{-4}$ shows that the gluon density
is large:
$$
x G (x,Q^2 ) \,\,\rightarrow \,\,40 - 50 \,\,{\rm at}
\,\,x_B\,=\,10^{-4}\,\,
{\rm and}\,\, Q^2\,=\,20\,GeV^2\,\,.
$$
The value for $xG(x,Q^2)\sim50$ was determined from the
$MRS D-'$ \cite{MRS} parameterization that roughly
describes the new data.
\par
2. The ZEUS collaboration measured the
diffraction dissociation (DD) \footnote{DD
is the process in which a number
of hadrons are produced in association with a recoil proton, whose
energy fraction is close to one.}
cross section and found
$$
\frac{\sigma^{DD}_{\gamma^*p}}{\sigma_{\gamma^*p}}
\,\,=\,\,\frac{5.2 \;nb}{80\; nb} \,\,=\,6 \times
\,10^{-2}\,{\rm at}\,x\,=\,10^{-4}\,\,{\rm and}\,\,Q^2\,>\,10\,GeV^2\,\,.
$$
where $\sigma_{\gamma^*p}$ is the total cross-section for
virtual-photon absorption.
We will say more about the diffractive component shortly.
\par
3. The structure function
$F_2$ can be described by the GLAP evolution equation
assuming that at the initial virtuality $Q^2_0 \sim 5 GeV^2$
 the gluon structure function increases rapidly at $ x \ra 0 $ as
$x G (x,Q^2_0) \,\propto \, x^{-\lambda}$ and $\lambda \geq 0.3$.
Note that at sufficiently large $Q_0^2$ $\lambda$ approaches
$\omega_0$, which occurs in the solution of the BFKL equation
(see section 11.2).
\par
 4. Both collaborations measured the total photoproduction
cross section $\s_t$ for
very small virtualities of the incoming photon and also for real photons.
This cross section exhibits a common property of soft
processes, namely $\sigma_t \ra   s^{\D}$ where $\D \sim 0.08$.
%
\subsection{The Value of the Shadowing Corrections from HERA Data}
Let us first try to understand the large number of gluons
(between 40 and 50 at $x=10^{-4}$)).
An interesting way to do
this is to compare the proton at $x = 10^{-4}$ and $Q^2=15 GeV^2$
with an iron nucleus, which has a similar number of {\it nucleon}
constituents.
If we calculate the packing fraction by assuming the size of the
constituents to be $R_h = 0.86\; fm$ for the nucleons and
$r_G = 1/Q$ for the gluons, we find that
the two packing factors are very similar
if one assumes the
size of the gluon disc to be $R_h/3$ (see section 4.2).
Thus, in this sense, we see
that a proton $x = 10^{-4}$ and $Q^2=15 GeV^2$ is very
similar to an iron nucleus.
\par
We now estimate the size of shadowing correction  $\Delta F_2$ for
the deep inelastic structure function.
Let us write it as
\beq
F_2 ( x, Q^2 ) \,\,=\,\,F^{GLAP}_2 ( x, Q^2 ) \,\,-\,\,\Delta F_2 ( x,
Q^2 )\,\,,
\eeq
where $ F^{GLAP}_2 $ is the solution of  the usual (GLAP) evolution equation.
A straightforward estimate gives
\beq
\frac{\Delta F_2}{F_2}\,\,\propto \,\,\alpha_s <r_G^2>\rho
\,\,\rightarrow
\,\,0.1\,\,\,{\rm for}\,\,Q^2\,=\,15 \,GeV^2 \,\,{\rm and}
\,\, x \,=\,10^{-4}\,\,.
\eeq
However, the ZEUS data on diffraction dissociation
let us better
estimate the value of SC. Indeed, Levin  \& Wuesthoff \cite{LEWU} showed
that
\beq
| \Delta F_2 |\,\,=\,\,F^{DD}_2,
\eeq
where $F^{DD}_2$ relates to $\sigma^{DD}_{\gamma^*p}$ in the
same way that $F_2$ relates to $\sigma_{\gamma^*p}$. In other words
$|\Delta \sigma_{\gamma^*p}/\sigma_{\gamma^*p} |
\,\,=\,\,\sigma^{DD}_{\gamma* p}/
\sigma_{\gamma^*p}$.
{}From the ZEUS data we can then conclude that
$| \Delta F_2 |/F_2\,\,>\,\, 6 \cdot 10^{-2} \,\,.$
Note that in this case we can  begin not only to
discuss whether there are SC, but even
whether we could predict this SC value in our theory.
\par The
details of the evolution equation for $F^{DD}_2$ were discussed in
Ref. \cite{LEWU}, in which Levin \& Wuesthoff suggested measuring
the sum $ F_2 \,+\,F^{DD}_2 $ in which all contributions of SC cancel.
One can use the GLAP equation for this sum, even at small $x$.
\subsection{Saturation of the Parton Density or Different Physics for ``Soft"
and ``Hard" Processes?}
The HERA data allow us to raise the question that is the title
of this subsection. Indeed, we can distinguish two regions in
$Q^2$ that have quite different energy behavior of the
photoproduction cross section. The experimental
facts that pertain to these regions are as follows:

{\bf 1.} $ Q^2\,\, \ll \,\,1 \, GeV^2$. ($a$) The total
cross section is essentially constant here (or increases slightly
with energy); ($b$) $\s^{DD}/\s_t \,\approx \,10\% $.
Thus in this kinematic region the photoproduction process
seems to be a typical soft process like a hadron-hadron
collision.

{\bf 2.} $ Q^2\,\, > \,\,5\, GeV^2 $. ($a$) The total
cross section increases rapidly  with energy ($x$)
$\s_{\gamma^*p} \,\propto \, x^{-\lambda} $; ($b$)
$\s^{DD}_{\gamma^*p}/\s_{\gamma^*p} \,\sim \,10\% $.
This is the typical deeply inelastic scattering process.
Both experimental facts arise from the
contribution of a so-called
hard Pomeron, which is a solution to the BFKL equation \cite{BFKL}
and leads to $\s_{\gamma^*p}
\,\propto \,x^{- \omega_0} $ with $\omega_0 > 0.4$ while
the smallness of the ratio  $\s^{DD}_{\gamma^*p}/\s_{\gamma^*p}$
has a natural explanation in the small values of SC.

What happens for intermediate value of photon
virtualities $ 1 GeV^2\, < \,Q^2 \,< \,5 GeV^2$? There are two
different scenarios for this kinematic region:

{\bf 1. Landshoff picture} At $Q^2 < 1 GeV^2$ all experimental data for
photoproduction as well as for other soft processes can be described
by  the exchange of the soft Pomeron, which is a standard Regge pole, with
intercept $\a_{P} \,=\,1\,+\,\eps$ ($\eps \,\sim \,0.08 \,\ll\,\omega_0)$
(see Ref.\cite{DL}). The small value of the ratio
$\s^{DD}/\s_t$ can then be
interpreted as an indication that the SC
are small and can be treated perturbatively.
In particular Donnachie and  Landshoff considered only two Pomeron
exchanges. In this scenario the region of deeply inelastic scattering has
different underlying physics described instead by
a hard Pomeron. The distinction
between the hard and soft regions seems quite arbitrary at present.
However, this approach is very simple and provides an elegant description of
all available experimental data on soft processes.

{\bf 2. Saturation of the parton density}
The second scenario is intimately related to the hypothesis that the parton
density saturates. In this scenario the hard Pomeron is responsible for
the behavior of the total cross-section in both kinematic regions, but
the small value of the ratio $ \s^{DD}/\s_t$ is interpreted differently
for large and small virtualities: For large virtualities it supports a
small value of SC, but at small ones we interpret it as an indication that
the SC become very large and lead to a black-disc constituent quark.
The greatest advantages of this scenario are the unified description
of both the diffractive and inclusive processes, and that its
solid theoretical background, namely
properties of the  hard Pomeron and the shadowing correction
in perturbative QCD.
It is worthwhile to mention that strong SC correction gives rise
sumalteneously the saturation of the parton density, smooth behavior of
the total cross section at high energy, a natural explanation of the
transition from steep energy behavior of the ``hard" Pomeron to smooth
 energy dependance of the ``soft" total cross section and small value of
the diffraction dissociation cross section in both kinematic regions (see
ref. \cite{LERY} for more information ).
However the data have not yet been described within this hypothesis.

These two scenarios result in different behavior
in the region of intermediate virtualities of photon. In the second scenario
we expect some transition region with smooth behavior of $\s_t$ vs
$Q^2$. A first attempt to extract this behavior from available experimental
data shows that such a transition region does not contradict such scenario
\cite{ALLM}, but it is still
too early to draw a definite conclusion.

\section{How to measure the high density event.}
Let me list here the main ideas  how to measure the new physics that
 we anticipate at high density system of partons:

{\bf 1.}  The probability of double parton interaction should be large
(of the order of the maximum value of the packing factor ) about 20\%.

The double parton interaction can be seen not only as cross section
 for production of two pair of hard jets with the same value of
rapidity,but also as a cross section of the inclusive production of hadrons
 in the window of rapidity
$y + \Delta y, y - \Delta y$ where y is the rapidity of a hard jet with
transverse momentum $p_t$ and $\Delta y \,=\,
\ln \frac{p_t}{p_0}$, where $p_0$ is the transverse momentum of
produced  hadron. It could be also seen as a long range correlation
 in rapidity between produced hard jet and produced hadron which
is not specially hard.

{\bf 2.} In the high density event we should see
 the Landau - Pomeranchuk suppression of the emission of gluons with
transverse momentum smaller that the typical momentum $q_0(x_B)$
 which can be found from the equation:
\beq
\frac{x G (q^2_0 (x),x)}{ q^2_0(x) \pi R^2} \,\,=\,\,(PF)_{max}\,\,.
\eeq
Such a suppression can be seen as the deviation from the
 factorization theorem for jets with $p_t \leq q_0 (x) $.

{\bf 3.}  . Decorrelation effect for jets with transverse momentum $ p_t$
of the order of $q_0$. The value of transverse momentum for such a jet is
 compensated not by one jet in the opposite direction but by a
number of jets with
average transverse momentum about $q_0$.

{\bf 4.} Polarization of produced hadrons allows us to measure the typical
transverse momentum in the process since in the region of pQCD polarization
should be equal to zero.

{\bf 5.} It is seen directly from eq.(9) that the saturation reaches
 in the system with small size at larger transverse momentum (smaller
 value of the gluon structure function). So this is why we have to create
 experimentally such compact system. We have
 three ideas how to confine the gluons in the disc of the small size:

{\bf A.} to find a carrier of partons with small size. Even
 hadron could be such a carrier if the hypothesis of constituent quarks
with small radius will be confirmed experimentally. However better to use
the virtual photon or Pomeron. The last is
not well theoretically defined object and what is Pomeron is one of the
 questions that the future experiment should answer.
However even available experimental information confirm the idea
that Pomeron's size is much smaller that hadron one and of the order
 of $ R_P \approx \sqrt{\alpha'_P } \sim 0.5 GeV^{-1} \sim 0.1 Fm$.
Thus hard diffraction with Pomeron can
give a good possibility to localize the parton system in small disc
 and to see high density phenomena in the most
clear way.

{\bf B.} to find the experiment (microscope) that can resolve the small
 part of the hadron and investigate it in detail.
This idea is realized in so called Mueller - Navalet process \cite{MUENAV} or
``hot spot" hunting \cite{HOTSPOT} (see Fig.34).

This process allows us to measure the small $\propto \frac{1}{p_{t2}}$ part
 of the hadron and using the two jets with sufficiently large transverse
 momentum as a trigger e can study the system with large parton density
 in many details.

{\bf C.}  Bjorken \cite{LRG} pointed out that the large rapidity gap ( LRG)
 processes can give us new way to look inside the high density parton system.
Indeed, due to intimate relation between inelastic processes and elastic one
coming from new reggeon-like approach the process with the LRG such as two high
transverse momentum jet production with rapidities $y_1$ and $y_2$ but without
any hadron with rapidities $ y_1 > y_h > y_2$ can be described as
 the exchange of
 ``hard" Pomeron (see Fig.35). The properties of the `` hard"  Pomeron exchange
is well known
 theoretically (see ref. \cite{BFKL} ) and can be checked experimentally.

{\bf 6.} One of the way to measure the high parton density  is to
select the event with large multiplicity of produced hadrons. More discussion
 on this subject you can find in the report of Fermilab Working Group on QCD
\cite{FERMI}
\newpage
\begin{Huge}
{\bf TOMORROW:}
\begin{center}
{\bf EFFECTIVE THEORY at HIGH ENERGY}\\
{\bf from NONPERTURBATIVE QCD}\\
\end{center}
\end{Huge}
\section{Summary.}
Let us summarize where we stand in the theoretical development at high energy
physics and what of the fundamental problems listed in section 9.3 have been
solved.

As we have discussed in section 9.5 we struggle with two major problems in
pQCD: ($a$) Our perturbative series have a natural small parameter $\as$, but
its
smallness is often compensated by large logs, which are different in different
 processes. In the case of the processes at high energies ( small $x$) the
real parameter of the perturbation series is $\as \ln (1/x)$ which is not
small.
($b$) The perturbation series is asymptotic because the coefficient $C_n$ in
 eq.(45) behaves as $n!$ at large $n$.

The first problem is mainly technical and has been solved by the BFKL equation.
However this equation carries with two problems, namely, the next
 order corrections have yet not been found and the solution of the BFKL
 equation is influenced by confinement region. Moreover, such confinement
corrections have not been localized in something like the matrix element
 of some operator.

The second problem is much more complicated one and it can be solved
 only if we understand better the nonperturbative origin  (renormalons,
 instantons,etc) of behaviour of coefficients $C_n$. We are only in the
 beginning of the systematic study of the nonperturbative effects in high
energy problem ( see ref. \cite{RENOR}).

The origin of the shadowing corrections is also nonperturbative one, but it
 is new kind of nonperturbative phenomena which closely related to the high
density of the partonic system. At the moment we can control them theoretically
only in the transition region between pQCD and hdQCD in Fig.19.

Coming back to the list of our hopes in section 9.5 we can conclude that we
have understood better the kinematic region in which we can trust pQCD.
\section{Nearest Future = Theory in the transition region.}
I firmly believe that in the nearest future :

{\bf 1.}

a theory for the transition region between pQCD and hdQCD ( see Fig.19) based
on
the BFKL Pomeron and Pomeron interactions will be completed.

{\bf 2.}

we will understand better the manifestation of the BFKL Pomeron and
 SC (shadowing corrections) and coming experiments from HERA and Tevatron
will help us to understand better in which kinematic region we can restrict
ourselves by new Reggeon Calculus based on the BFKL Pomeron.

{\bf 3.}

we will find the values of all nonperturbative parameters that we need (like
$R$ in the GLR equation) with help of the nonperturbative QCD methods such as
lattice calculation or QCD sum rules.

The Mueller approach as well as the deep understanding of the correct degrees
of freedom allows us to start  to built the effective theory for high energy
 interaction. It is interesting to emphasize that this effective theory will
be a theory of interaction of the dimensional objects ( colour dipoles).
 Therefore, at high energy we have such a system of partons (particle) in
which we can study the theory of interaction of the dimension degrees of
freedom. It seems very attractive, because we can check our theoretical
 imagination with experimental data, while the string gang who is doing
 basically the same trying to understand how can we build the relativistic
 theory
for dimension degrees of freedom  still have to rely only on theory without
any experimental support.
\section{Toward the effective theory.}
I hope that I have  convinced you that  the main direction of our
movement is toward the effective theory based on nonperturbative QCD approach,
 because the principle problems of high density partonic system  cannot be
 solved without understanding of npQCD.  We hope that in the future we will
find:

1. a selfconsistent effective theory for a high density system in QCD;

2.the analytic solution of hdQCD,using the smallness of the coupling
 QCD constant in this kinematic region;

3. a new collective phenomena in hdQCD.

At the present we can see two harbingers of this shining future:

\subsection{ Effective Lagrangian for high energy QCD.}
 The first  attempt to develop nonperturbative approach to low $x_B$
 problems was to write down the effective Lagrangian
for the region of small $ x_B$.
Intrinsically we assume that such a Lagrangian
should be simpler that the QCD one and allows one to apply some direct
 numerical procedure (lattice calculation for example) to calculate the
scattering amplitude with this Lagrangian. It should be stressed the attempts
to calculate the amplitude with full QCD Lagrangian  have failed by now.
At the moment we have two effective theories on the market: one was proposed
 by Lipatov \cite{LIPET} which looks not much simpler that  full
QCD
 but it certainly incorporates all results of perturbative calculations,
 and the second was suggested by Verlinde and Verlinde \cite{VVET}, which
 is much simpler and is suited for lattice -like calculation but it has not
been checked how well this effective theory describes the perturbative results.
Moreover there is some indication \cite{TAN} that this theory cannot describe
the virtual correction in the BFKL Pomeron.

\subsection{ Thermodynamics of high density QCD.}
I firmly believe that we need to write down the correct kinetic equation for
high density QCD. Such an approach has certainly at least one big advantage:
the smooth matching with the GLR equation. Unfortunately we have not yet
 understood  how  to write such an equation in our nonequilibrium
 situation. However we have understood better the physical meaning of the
new typical momentum in our parton cascade ($ \langle |p_t | \rangle $
in section 11.2 ). It turns out that this momentum is the Landau - Pomeranchuk
 momentum for our parton medium \cite{LL}. Thus the gluon emission with
transverse momentum less that $\langle | p_t | \rangle $ is small due to
 destructive interference between emission before and after collision of the
parton with other partons in the medium. The experience with the calculation of
the anomalous dimension of high twist gluonic operators also give us
 understanding why for bosonic degrees of freedom such as a gluon there could
be
saturation of gluon density. Indeed, our cascade  is rather  a one dimensional
 one and for such a system the direction of motion plays the role of spin for
fermions.
\section{Conclusions.}
We can perhaps summarize our discussion with the following statements.

1. The new kinematic region of {\it high density QCD} is
wide open for theoretical and experimental investigation.

2. Perturbative QCD methods have been developed to explore
the transition region between the region routinely
analyzed by perturbative calculations
for hard processes  and the hdQCD region.

3. Between the kinematic region of soft processes
(key words: unitarity, analyticity, Reggeons, Reggeon Calculus, hadronization
 models, etc) and the kinematical region of hard processes (key words:
 quarks, gluons, parton distributions, asymptotic freedom, pQCD, evolution
 equations, etc) is the region of
{\bf semi-hard
processes},  where the
cross sections can be as large as
the geometrical
size of the hadron but with a typical transverse momentum
large enough that methods of pQCD can be applied.

4.  New probes have been developed for this high-density QCD region.
Although we still lack some theoretical
understanding and cannot provide
reliable estimates for these probes, we
firmly believe that they provide a deeper insight into
QCD and could lead to new experimental discoveries.

{\bf Phase Transition.}

We can represent the predictions of the previous section as a phase
transition between different forms of hadronic (partonic) matter.
Our prediction can then be summarized as the possibility of a
phase transition from a {\it quark - gluon ideal gas} to a { \it gluon liquid}
(see Fig. 36).

Indeed, in the {\it pQCD} region
(see Fig.19) we can regard a partonic system as an ideal gas of
quarks and gluons because we can neglect their interactions (only
emission is essential here).

When the density of partons increases we approach
the region where we must consider the interactions between
partons in a simple approximation and in a situation where this interaction is
not too big. This then is a
Van der Waals gas of quarks and gluons. Here we have
a  theory based on a nonlinear evolution equation that is the
analogue of the
Van der Waals equation for a parton system.

For even larger values of the density,
the smallness of the coupling constant
results a small correlation between partons. We
call this system of partons a {\it gluon liquid}.
We recall that the density is high primarily because of
emission of gluons and
quarks inside this system, which still can be treated in a perturbative way.

The border between the gluon liquid and the Van der Waals gas is
the critical line we
discussed in Section 12.3.
The trajectory of the GLAP evolution equation that touches the critical line
separates the Van der Waals gas from the ideal gas.
The asymptotic of these borders are also calculable.

In conclusion we want also to recommend you the list of the reviews on the
 subject \cite{REV} which, we hope, you will be able to understand after
these lectures.

\section*{Acknowledgments:}
I would like to thank  the CNPq for financial support and the organizing
 committee of the Third Gleb Watagin school for invitation to read these
 lectures.  I am very grateful to all my friends
 especially F. Caruso, A. Gotsman, U.Maor and A. Santoro
 with whom I discussed the
ideas of such lectures and who supported me in this rather difficult job.

\newpage
\begin{Large}
\begin{center}
{\bf Figure Caption:}
\end{center}
\end{Large}
\begin{tabular}{r l}
{\bf Fig.1:}& The exchange of the resonance with spin $j$.\\
& \\
{\bf Fig.2:}& The exchange of the Reggeon.\\
 & \\
{\bf Fig.3:}& The Reggeon trajectories.\\
 & \\
{\bf Fig.4:}& Resonance - Reggeon Duality.\\
 & \\
{\bf Fig.5:}& The Pomeron structure in the Veneziano model.\\
 & \\
{\bf Fig.6:}&The structure of the parton cascade.\\
 & \\
{\bf Fig.7:}& The random walk of parton in transverse plane.\\
 & \\
{\bf Fig.8:}&The single diffraction dissociation process in the Pomeron
 approach\\
 & \\
{\bf Fig.9:}& The triple Pomeron diagram.\\
 & \\
{\bf Fig.10:}& An example of enchanced Pomeron diagrams.\\
 & \\
{\bf Fig.11} & The AGK cutting rules for two Pomeron exchange.\\
 & \\
{\bf Fig.12:}& Hadron -dueteron interaction.\\
 & \\
{\bf Fig.13:}& The process with multiplicity $2 <n_N> $ ($a$) and $ <n _N>$
 ($b$).\\
 & \\
{\bf Fig.14:}& The  $\nu$ Pomeron exchange diagram with $\mu$ cut Pomeron.\\
 & \\
{\bf Fig.15:}& Pomeron diagrams for the total,elastic and diffraction
dissociation cross sections.\\
 & \\
{\bf Fig.16:}& Pomeron diagrams for inclusive cross section ($a$) and rapidity
correlations ($b$).\\
 & \\
{\bf Fig.17:}& One and two parton shower processes.\\
\end{tabular}
\newpage
\begin{tabular} {r l}
{\bf Fig.18:}& Single diffraction dissociation in Reggeon approach.\\
 & \\
{\bf Fig.19:}& The map of QCD.$\rho$ is the parton (gluon ) density\\
&and $r$ is
the distances resolved in an experiment.\\
 & \\
{\bf Fig.20:}& The ``Pomeron" Calculus in QCD.\\
 & \\
{\bf Fig.21:} &The Born Approximation (BA) of pQCD.\\
 & \\
{\bf Fig.22:}& The next to BA of pQCD for quark-quark scattering:\\
 & emission of
 one extra gluon.\\
& \\
{\bf Fig.23:}& The next to BA of pQCD for quark-quark scattering:\\
& $\as^2$
 correction to elastic amplitude.\\
& \\
{\bf Fig.24:}& The next to BA of pQCD for quark -quark scattering:\\
& gauge
 invariance trick.\\
 & \\
{\bf Fig.25:}& The next to BA of pQCD for quark -quark scattering:\\
& the resulting
answer for one extra gluon emission.\\
 & \\
{\bf Fig.26:}& The relation between colour coefficients.\\
 & \\
{\bf Fig.27:}& The BFKL equation.\\
 & \\
{\bf Fig.28:}& The basic branching process in DIS.\\
 & \\
{\bf Fig.29:}&The parton distribution in the transverse plane.\\
 & \\
{\bf Fig.30:}& Structure of the parton cascades at low $x$ and\\
 & the coherence in ``ladder" diagrams.\\
\end{tabular}
\newpage
\begin{tabular} {r l}
{\bf Fig.31:}& The random walk picture in $\ln k^2_t$.\\
 & \\
{\bf Fig.32:}& The behaviour of $xG(x,Q^2)$ versus $Q^2$ at fixed $x$.\\
 & \\
{\bf Fig.33:}& Trajectory of the nonlinear GLR equation.\\
 & \\
{\bf Fig.34:}& The Mueller - Navalet process.\\
 & \\
{\bf Fig.35:}& Large Rapidity Gap process.\\
 & \\
{\bf Fig.36:}& The phase diagram for hadron - hadron collisions for zero
 rapidity\\
 & in cm frame. W is the absorption probability defined in eq. (103).\\
\end{tabular}
\end{document}